\documentclass[aps,prl,twocolumn,groupedaddress]{revtex4}

\usepackage{array,hhline,dcolumn} 
\usepackage{rotating} 

\newcommand{\gtwid}{\mathrel{\raise.3ex\hbox{$>$\kern-.75em\lower1ex\hbox{$\sim$}}}}
\newcommand{\ltwid}{\mathrel{\raise.3ex\hbox{$<$\kern-.75em\lower1ex\hbox{$\sim$}}}}

\begin{document}
%
\title{Significant Excess of Electronlike Events in the MiniBooNE Short-Baseline 
Neutrino Experiment}

\author{
        A.~A. Aguilar-Arevalo$^{13}$,
        B.~C.~Brown$^{6}$, L.~Bugel$^{12}$,
        G.~Cheng$^{5}$, 
	J.~M.~Conrad$^{12}$,
        R.~L.~Cooper$^{10,15}$, R.~Dharmapalan$^{1,2}$, A.~Diaz$^{12}$,
        Z.~Djurcic$^{2}$, D.~A.~Finley$^{6}$, R.~Ford$^{6}$,
        F.~G.~Garcia$^{6}$, G.~T.~Garvey$^{10}$,
        J.~Grange$^{7}$, E.-C.~Huang$^{10}$,
        W.~Huelsnitz$^{10}$, C.~Ignarra$^{12}$, 
        R.~A. ~Johnson$^{3}$, G.~Karagiorgi$^{5}$, T.~Katori$^{12,16}$,
        T.~Kobilarcik$^{6}$,
        W.~C.~Louis$^{10}$, C.~Mariani$^{19}$, W.~Marsh$^{6}$,
        G.~B.~Mills$^{10,\dagger}$,
        J.~Mirabal$^{10}$, J.~Monroe$^{18}$,
        C.~D.~Moore$^{6}$, J.~Mousseau$^{14}$,
        P.~Nienaber$^{17}$, J.~Nowak$^{9}$,
        B.~Osmanov$^{7}$, Z.~Pavlovic$^{6}$, D.~Perevalov$^{6}$,
	H.~Ray$^{7}$, B.~P.~Roe$^{14}$,
        A.~D.~Russell$^{6}$,
        M.~H.~Shaevitz$^{5}$,
        J.~Spitz$^{14}$, I.~Stancu$^{1}$,
        R.~Tayloe$^{8}$, R.~T.~Thornton$^{10}$, M.~Tzanov$^{4,11}$, 
	R.~G.~Van~de~Water$^{10}$,
        D.~H.~White$^{10,\dagger}$, D.~A.~Wickremasinghe$^{3}$, 
        E.~D.~Zimmerman$^{4}$ \\
\smallskip
(The MiniBooNE Collaboration)
\smallskip
}
\smallskip
\smallskip
\affiliation{
$^1$University of Alabama; Tuscaloosa, AL 35487, USA \\
$^2$Argonne National Laboratory; Argonne, IL 60439, USA \\
$^3$University of Cincinnati; Cincinnati, OH, 45221, USA \\
$^4$University of Colorado; Boulder, CO 80309, USA \\
$^5$Columbia University; New York, NY 10027, USA \\
$^6$Fermi National Accelerator Laboratory; Batavia, IL 60510, USA \\
$^7$University of Florida; Gainesville, FL 32611, USA \\
$^8$Indiana University; Bloomington, IN 47405, USA \\
$^9$Lancaster University; Lancaster LA1 4YB, UK \\
$^{10}$Los Alamos National Laboratory; Los Alamos, NM 87545, USA \\
$^{11}$Louisiana State University; Baton Rouge, LA 70803, USA \\
$^{12}$Massachusetts Institute of Technology; Cambridge, MA 02139, USA \\
$^{13}$Instituto de Ciencias Nucleares; Universidad Nacional Aut\'onoma de M\'exico; CDMX 04510, M\'exico \\
$^{14}$University of Michigan; Ann Arbor, MI 48109, USA \\
$^{15}$New Mexico State University; Las Cruces, NM 88003, USA \\
$^{16}$Queen Mary University of London; London E1 4NS, UK \\
$^{17}$Saint Mary's University of Minnesota; Winona, MN 55987, USA \\
$^{18}$Royal Holloway, University of London; Egham TW20 0EX, UK \\
$^{19}$Center for Neutrino Physics; Virginia Tech; Blacksburg, VA 24061, USA \\
$^\dagger$Deceased \\
}

\date{\today}

\begin{abstract}
The MiniBooNE experiment at Fermilab reports results from an 
analysis of  $\nu_e$ appearance data from $12.84 \times 10^{20}$ protons
on target in neutrino mode, an increase of approximately a factor of two over
previously reported results.  A $\nu_e$ charged-current quasielastic event 
excess of $381.2 \pm 85.2$ events ($4.5 \sigma$)
is observed in the energy range $200<E_\nu^{QE}<1250$~MeV.  Combining these data
with the $\bar \nu_e$ appearance data from $11.27 \times 10^{20}$ protons
on target in antineutrino mode, a total $\nu_e$ plus $\bar \nu_e$ charged-current quasielastic
event excess of $460.5 \pm 99.0$ events ($4.7 \sigma$)
is observed. If interpreted in a 
two-neutrino oscillation model, ${\nu}_{\mu} \rightarrow {\nu}_e$, the best oscillation
fit to the excess has a probability of $21.1\%$, while
the background-only fit has a $\chi^2$ probability of $6 \times 10^{-7}$ relative to the best
fit.  The MiniBooNE data are consistent in energy and magnitude
with the excess of events reported by the Liquid Scintillator 
Neutrino Detector (LSND), 
and the significance of the combined LSND and MiniBooNE excesses is $6.0 \sigma$.
A two-neutrino oscillation interpretation of the data would require at least four neutrino types 
and indicate physics beyond the three neutrino paradigm.
Although the data are fit with a two-neutrino oscillation model,
other models may provide
better fits to the data.
\end{abstract}


\maketitle



Evidence for short-baseline neutrino anomalies at an $L/E_\nu \sim 1$ m/MeV, where $E_\nu$ 
is the neutrino energy and $L$ is the distance that the neutrino traveled before detection, comes 
from both neutrino appearance and disappearance experiments. 
The appearance anomalies include the excess of $\nu_e$ and $\bar \nu_e$ charge-current quasielastic
(CCQE) events observed by the 
LSND \cite{lsnd} and MiniBooNE \cite{mb_osc,mb_oscnew}
experiments, while the disappearance anomalies, although not completely consistent,
include the deficit of $\nu_e$ and $\bar \nu_e$
events observed by reactor \cite{reactor}
and radioactive-source experiments \cite{radioactive}. 
As the masses and mixings within the 3-generation neutrino matrix have been 
attached to solar and long-baseline neutrino experiments, more exotic models 
are typically used to explain these anomalies, including,
for example, 3+N neutrino oscillation models 
involving three active neutrinos and N additional sterile neutrinos 
\cite{sorel,karagiorgi,collin,giunti,giunti2,kopp,kopp2,white_paper,3+2}, 
resonant neutrino oscillations \cite{resonant}, Lorentz violation 
\cite{lorentz}, sterile neutrino decay \cite{sterile_decay}, sterile neutrino nonstandard interactions
\cite{NSI}, and sterile neutrino extra dimensions
\cite{sterile_extra}.
This Letter presents improved MiniBooNE $\nu_e$ and $\bar \nu_e$ appearance 
results, assuming two-neutrino oscillations with probability $P=\sin^2(2\theta) 
\sin^2 (1.27 \Delta m^2 L/E)$, where $\theta$ is the mixing angle, $\Delta m^2$ (eV$^2$/c$^4$)
is the difference in neutrino mass eigenstates squared, $L$ (m)
is the distance traveled by the neutrino, and $E$ (MeV) is the neutrino energy.


The booster neutrino beam (BNB) at Fermilab delivers to the MiniBooNE
experiment a flux of neutrinos and antineutrinos that is simulated
using information from external measurements \cite{mb_flux}. The BNB is produced by 8 GeV protons from 
the Fermilab booster interacting on
a beryllium target inside a magnetic focusing horn. Depending on the polarity of the horn,
either $\pi^+$ are focused and $\pi^-$ are defocused to produce a fairly pure beam of $\nu_{\mu}$,
or $\pi^-$ are focused and $\pi^+$ are defocused to produce a somewhat pure beam of $\bar \nu_{\mu}$.
In neutrino mode, the $\nu_\mu$, $\bar \nu_\mu$, $\nu_e$, and $\bar \nu_e$ flux
contributions at the detector are 93.5\%, 5.9\%, 0.5\%, and 0.1\%, respectively, while
in antineutrino mode, the flux
contributions are 15.7\%, 83.7\%, 0.2\%, and 0.4\%, respectively.
The $\nu_\mu$ and $\bar{\nu}_{\mu}$ fluxes peak at approximately 600 MeV and 400 MeV, respectively. 

The MiniBooNE detector is described in detail in reference \cite{mb_detector}. 
The detector consists of
a 12.2 m diameter sphere filled with 818 tonnes of pure mineral oil (CH$_{2}$) and
is located 541 m from the beryllium target. The detector is covered by 1520 8-inch
photomultiplier tubes (PMTs), where 1280 PMTs are in the interior detector region and 240
PMTs are located in the optically isolated outer veto region. Charged particles produced
by neutrino interactions in the mineral oil emit both directed Cherenkov light and
isotropic scintillation light that is detected by the PMTs.
Event reconstruction \cite{mb_recon} and particle identification make use of
the hit PMT charge and time information, and the reconstructed neutrino energy, $E_\nu^{QE}$, 
is estimated from the measured energy and angle of the outgoing muon 
or electron, assuming the kinematics of CCQE scattering \cite{ccqe}.

From 2002-2017, the MiniBooNE experiment has collected a total of $11.27 \times 10^{20}$ 
protons on target (POT) in antineutrino mode, $12.84 \times 10^{20}$ POT
in neutrino mode, and a further $1.86 \times 10^{20}$ POT in a special
beam-off target mode to search for sub-GeV dark matter \cite{mb_dm}. 
The neutrino sample has approximately doubled in size since the
previous publication \cite{mb_oscnew}.  
The published neutrino-mode data correspond to $6.46 \times 10^{20}$ POT,
while $6.38 \times 10^{20}$ POT were obtained in 2016 and 2017.
During the 15 years of
running, the BNB and MiniBooNE detector have been stable to within
2\% in neutrino energy.


The analysis is optimized to measure $\nu_e$ and $\bar \nu_e$ induced CCQE 
events, and the event reconstruction \cite{mb_recon} and selection are identical to the previous
analysis \cite{mb_oscnew}. The average selection efficiency is
$\sim 20\%$ ($\sim 0.1\%$) for $\nu_e$-induced CCQE events ($\nu_\mu$-induced background events)
generated over the fiducial volume.
The fraction of CCQE events in antineutrino mode that
are from wrong-sign neutrino events was determined from the angular 
distributions of muons created in CCQE interactions and
by measuring CC single $\pi^+$ events \cite{wrong_sign}.

The predicted but unconstrained $\nu_e$ and $\bar{\nu}_e$ CCQE background events for 
the neutrino energy range $200<E_\nu^{QE}<1250$~MeV 
are shown in Table \ref{signal_bkgd} for both neutrino mode and antineutrino mode \cite{ryan}. 
See appendices for more information on backgrounds.
The upper limit of 1250 MeV corresponded to a small value of L/E and
was chosen by the collaboration before unblinding the data in 2007. 
The lower limit of 200 MeV is chosen because we constrain the $\nu_e$ events with the 
CCQE $\nu_\mu$ events 
and our CCQE $\nu_\mu$ event sample only goes down to 200 MeV, as we require a visible Cherenkov ring from the muon.
The estimated sizes of the intrinsic $\nu_e$ and gamma backgrounds 
are based on MiniBooNE event measurements and uncertainties from these constraints 
are included in the analysis.  
The intrinsic $\nu_e/\bar\nu_e$ background from muon decay is
directly related to the large sample of 
observed $\nu_\mu / \bar\nu_\mu$ events, as these events constrain the muons 
that decay in the 50 m decay region.  
This constraint uses a joint fit of
the observed $\nu_\mu / \bar\nu_\mu$ and $\nu_e/\bar\nu_e$ events, assuming 
that there are no substantial $\nu_\mu / \bar\nu_\mu$ disappearance oscillations.  The other
intrinsic $\nu_e$ background component, from kaon decay, is constrained by fits to kaon production
data and SciBooNE measurements \cite{sciboone_kaon}.
The intrinsic $\nu_e$ background from pion decay ($1.2 \times 10^{-4}$ branching ratio) and hyperon decay are very small. 
Other backgrounds from mis-identified $\nu_{\mu}$ or $\bar{\nu}_{\mu}$ 
\cite{mb_numuccqe,mb_numuccpi} events are also constrained by the observed CCQE sample.

The gamma background from neutral-current (NC) $\pi^0$ production and
$\Delta \rightarrow N\gamma$ radiative
decay \cite{hill_zhang,nieves} are constrained by the associated large two-gamma sample (mainly from $\Delta$ 
production) observed in the MiniBooNE data, where 
$\pi^0$ measurements \cite{mb_pi0} are used to constrain the $\pi^0$ background. The $\pi^0$
background measured in the first and second neutrino data sets were found to be consistent, resulting in a
lower statistical background uncertainty for the combined data.
Other neutrino-induced single gamma 
production processes are included in the theoretical predictions, which agree well with
the MiniBooNE estimates \cite{hill_zhang,othergamma}. 
Single-gamma backgrounds
from external neutrino interactions (``dirt" backgrounds) are estimated using topological and spatial cuts
to isolate the events whose vertices are near the edge of the detector and
point towards the detector center \cite{mb_lowe}.  With the larger data set, 
the background from external neutrino
interactions is now better determined to be approximately 7\% larger, but with smaller
uncertainty, than in the previous
publication \cite{mb_oscnew}. A new technique to
measure or constrain the gamma and dirt backgrounds based on event timing relative to
the beam is in development.

\begin{table}[t]
\vspace{-0.1in}
\caption{\label{signal_bkgd} \em The expected (unconstrained) number of events
for the $200<E_\nu^{QE}<1250$~MeV neutrino 
energy range from all of the backgrounds in the $\nu_e$ and $\bar{\nu}_e$ 
appearance analysis before using the constraint from the CC $\nu_\mu$ events. 
Also shown are the constrained background, as well as
the expected number of events corresponding to the LSND best fit 
oscillation probability of 0.26\%, assuming oscillations at large $\Delta m^2$.
The table shows
the diagonal-element systematic plus statistical uncertainties, which become
substantially reduced in the oscillation fits when correlations
between energy bins and between the electron and muon neutrino events
are included. The antineutrino numbers are from a previous analysis \cite{mb_oscnew}.
}
\small
\begin{ruledtabular}
\begin{tabular}{ccc}
Process&Neutrino Mode&Antineutrino Mode \\
\hline
$\nu_\mu$ \& $\bar \nu_\mu$ CCQE & 73.7 $\pm$ 19.3 & 12.9 $\pm$ 4.3 \\
NC $\pi^0$ & 501.5 $\pm$ 65.4 & 112.3 $\pm$ 11.5 \\
NC $\Delta \rightarrow N \gamma$ & 172.5  $\pm$ 24.1 & 34.7 $\pm$ 5.4 \\
External Events & 75.2 $\pm$ 10.9 & 15.3 $\pm$ 2.8 \\
Other $\nu_\mu$ \& $\bar \nu_\mu$ & 89.6 $\pm$ 22.9 & 22.3 $\pm$ 3.5 \\
\hline
$\nu_e$ \& $\bar \nu_e$ from $\mu^{\pm}$ Decay & 425.3 $\pm$ 100.2 & 91.4 $\pm$ 27.6 \\
$\nu_e$ \& $\bar \nu_e$ from $K^{\pm}$ Decay & 192.2  $\pm$ 41.9 & 51.2 $\pm$ 11.0 \\
$\nu_e$ \& $\bar \nu_e$ from $K^0_L$ Decay & 54.5 $\pm$ 20.5 & 51.4 $\pm$ 18.0 \\
Other $\nu_e$ \& $\bar \nu_e$ & 6.0 $\pm$ 3.2 & 6.7 $\pm$ 6.0 \\
\hline
Unconstrained Bkgd. & $1590.6 \pm 176.9$ & $398.2  \pm 49.7$ \\
Constrained Bkgd. & $1577.8 \pm 85.2$ &  $398.7 \pm 28.6$ \\
\hline
Total Data & 1959 &478 \\
Excess & 381.2 $\pm$ 85.2 & 79.3 $\pm$ 28.6 \\
\hline
0.26\% (LSND) $\nu_\mu \rightarrow \nu_e$ & 463.1 & 100.0 \\
\end{tabular}
\vspace{-0.2in}
\end{ruledtabular}
\normalsize
\end{table}

Systematic uncertainties are determined by considering the predicted
effects on the $\nu_\mu$, $\bar{\nu}_{\mu}$, $\nu_e$, and $\bar{\nu}_e$ CCQE rates 
from variations of uncertainty parameters.
The parameters include uncertainties in the neutrino and antineutrino flux estimates, 
uncertainties in neutrino cross sections, most of which are determined by 
in situ cross-section 
measurements at MiniBooNE \cite{mb_numuccqe,mb_pi0}, uncertainties from nuclear effects, and uncertainties in 
detector modeling and reconstruction. 
A covariance matrix in bins of $E^{QE}_{\nu}$ is constructed 
by considering the variation from each source of systematic uncertainty on the $\nu_e$ and $\bar{\nu}_e$ CCQE signal and background, and the
$\nu_\mu$ and $\bar{\nu}_{\mu}$ CCQE prediction as a function of $E_{\nu}^{QE}$.
This matrix includes correlations between any of the $\nu_e$ and $\bar{\nu}_e$ CCQE signal and background and 
$\nu_\mu$ and $\bar{\nu}_{\mu}$ CCQE samples, and is used in the $\chi^2$ calculation of the oscillation fits.

Table \ref{signal_bkgd} also shows the expected number of events corresponding to the LSND best fit
oscillation probability of 0.26\%, assuming oscillations at large $\Delta m^2$.
LSND and MiniBooNE have the same average value of L/E, but MiniBooNE has a larger range of L/E. Therefore, the appearance 
probabilities for LSND and MiniBooNE should not be exactly the same at lower L/E values.

\begin{figure}[tbp]
\vspace{+0.1in}
\centerline{\includegraphics[angle=0, width=9.0cm]{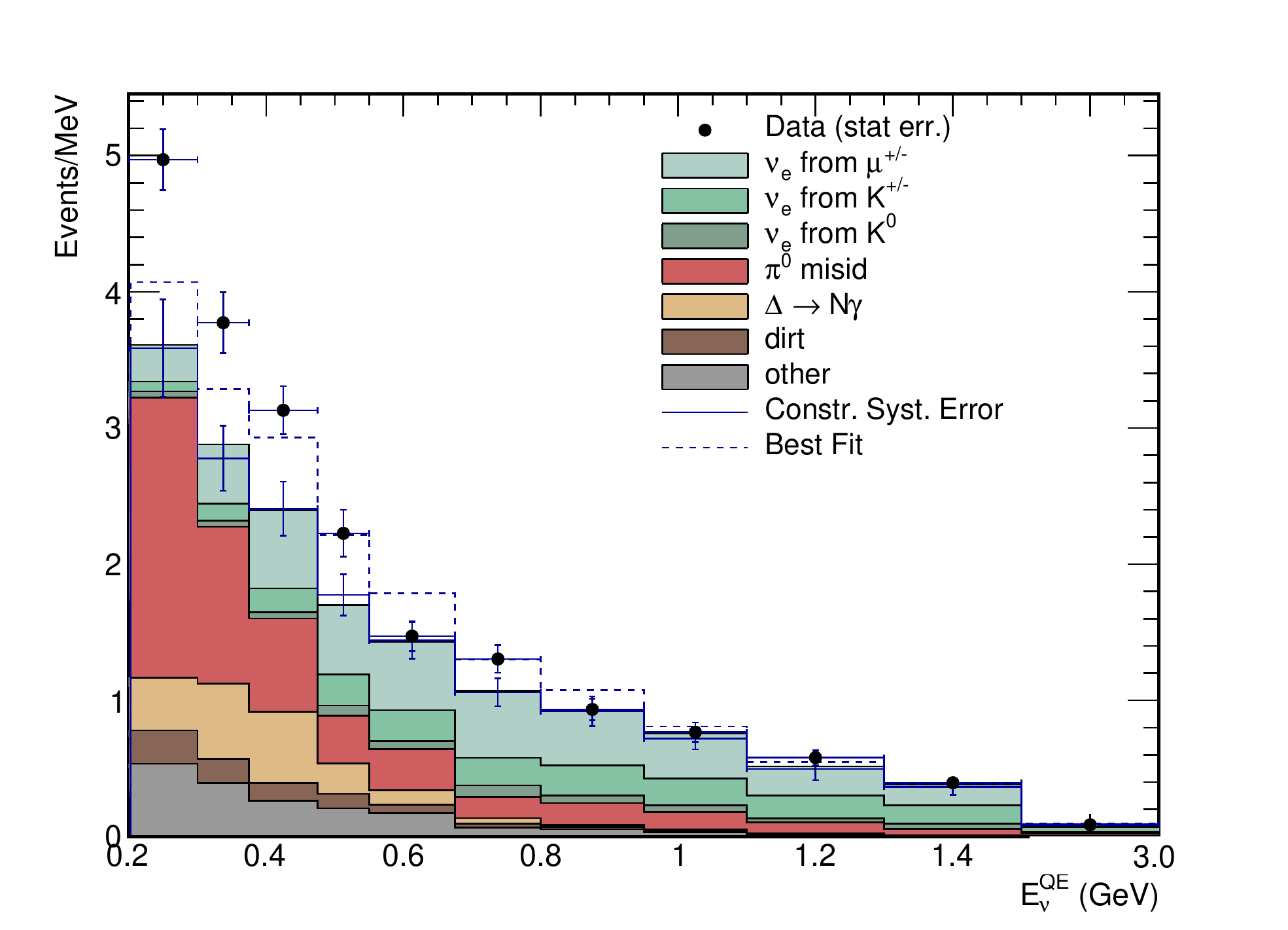}}
\vspace{-0.2in}
\caption{The MiniBooNE neutrino mode 
$E_\nu^{QE}$ distributions, corresponding to the total $12.84 \times 10^{20}$ POT data, 
for ${\nu}_e$ CCQE data (points with statistical errors) and background 
(histogram with systematic errors). The dashed curve shows 
the best fit to the neutrino-mode data assuming two-neutrino oscillations.
The last bin is for the energy interval from 1500-3000 MeV.} 
\label{excessnat}
\vspace{0.1in}
\end{figure}
\begin{figure}[tbp]
\vspace{-0.0in}
\centerline{\includegraphics[angle=0, width=9.0cm]{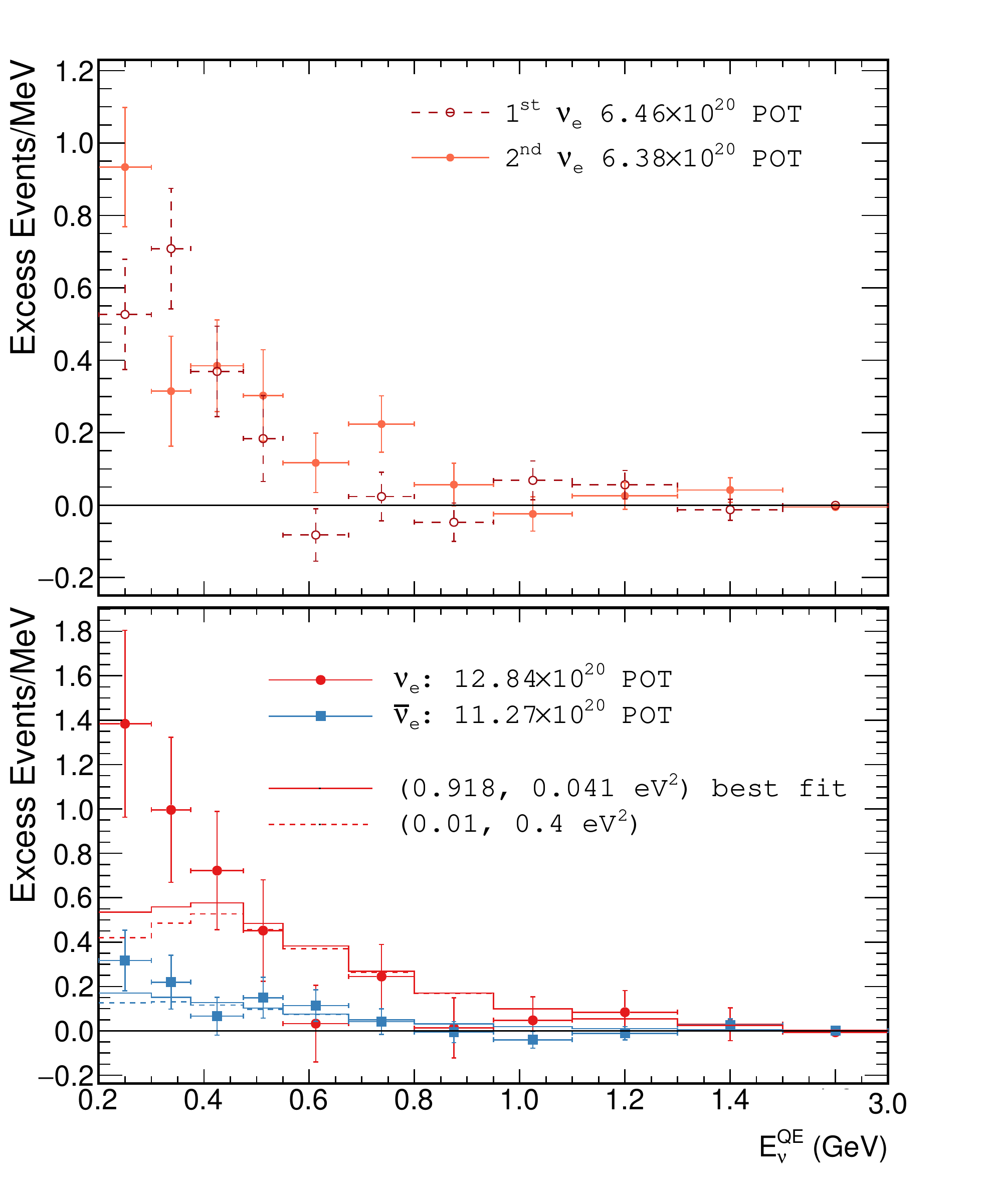}}
\vspace{-0.2in}
\caption{The top plot shows the MiniBooNE event excesses in neutrino mode as a function of $E_\nu^{QE}$
from the first
$6.46 \times 10^{20}$ POT data and the second $6.38 \times 10^{20}$ POT data.
The bottom plot shows the total event excesses 
in both neutrino mode and antineutrino mode, corresponding to $12.84 \times 10^{20}$ POT 
and $11.27 \times 10^{20}$ POT, respectively.
The solid (dashed) curve is the best fit ($1 \sigma$ fit point) to the neutrino-mode and antineutrino-mode data
assuming two-neutrino oscillations. The last bin is for the energy interval from 1500-3000 MeV.
Error bars include only statistical uncertainties for the top plot and 
both statistical and correlated systematic uncertainties for the bottom plot.}
\label{Excess_old_new}
\vspace{0.1in}
\end{figure}

Fig. \ref{excessnat} shows the $E_\nu^{QE}$ distribution for 
${\nu}_e$ CCQE data and background in
neutrino mode for the total $12.84 \times 10^{20}$ POT data. 
Each bin of reconstructed $E_\nu^{QE}$
corresponds to a distribution of ``true'' generated neutrino energies,
which can overlap adjacent bins.
In neutrino mode, a total of 1959 data events pass
the $\nu_e$ CCQE event selection requirements with $200<E_\nu^{QE}<1250$~MeV,
compared to a background expectation of $1577.8 \pm 39.7 (stat.) \pm 75.4 (syst.)$ events.
The excess is then $381.2 \pm 85.2$ events or a $4.5 \sigma$ effect.
Note that the 162.0 event excess in the first $6.46 \times 10^{20}$ POT data 
is approximately $1 \sigma$ lower than the average excess, while the
219.2 event excess in the second $6.38 \times 10^{20}$ POT data 
is approximately $1 \sigma$ higher than the average excess.
Fig. \ref{Excess_old_new} shows the excess events in neutrino mode from the first
$6.46 \times 10^{20}$ POT data and the second $6.38 \times 10^{20}$ POT data
(top plot). 
Combining the MiniBooNE neutrino and antineutrino data, there are a total of 2437 events 
in the $200<E_\nu^{QE}<1250$~MeV energy region, 
compared to a background expectation of $1976.5 \pm 44.5(stat.) \pm 88.5(syst.)$ events. 
This corresponds to a total $\nu_e$ plus $\bar \nu_e$ CCQE excess of $460.5 \pm 99.0$ events
with respect to expectation or a $4.7 \sigma$ excess. 
Fig. \ref{Excess_old_new} (bottom plot) shows the total event excesses
as a function of $E_\nu^{QE}$ in both neutrino mode and antineutrino mode.
The dashed curves show the two-neutrino oscillation predictions at the best-fit point ($\Delta m^2=0.041$ eV$^2$, 
$\sin^22\theta = 0.92$), as well as at a point within $1 \sigma$ of the best-fit point ($\Delta m^2=0.4$ eV$^2$, 
$\sin^22\theta = 0.01$).

\begin{figure}[tbp]
\vspace{-0.25in}
 \centerline{\includegraphics[width=9.0cm]{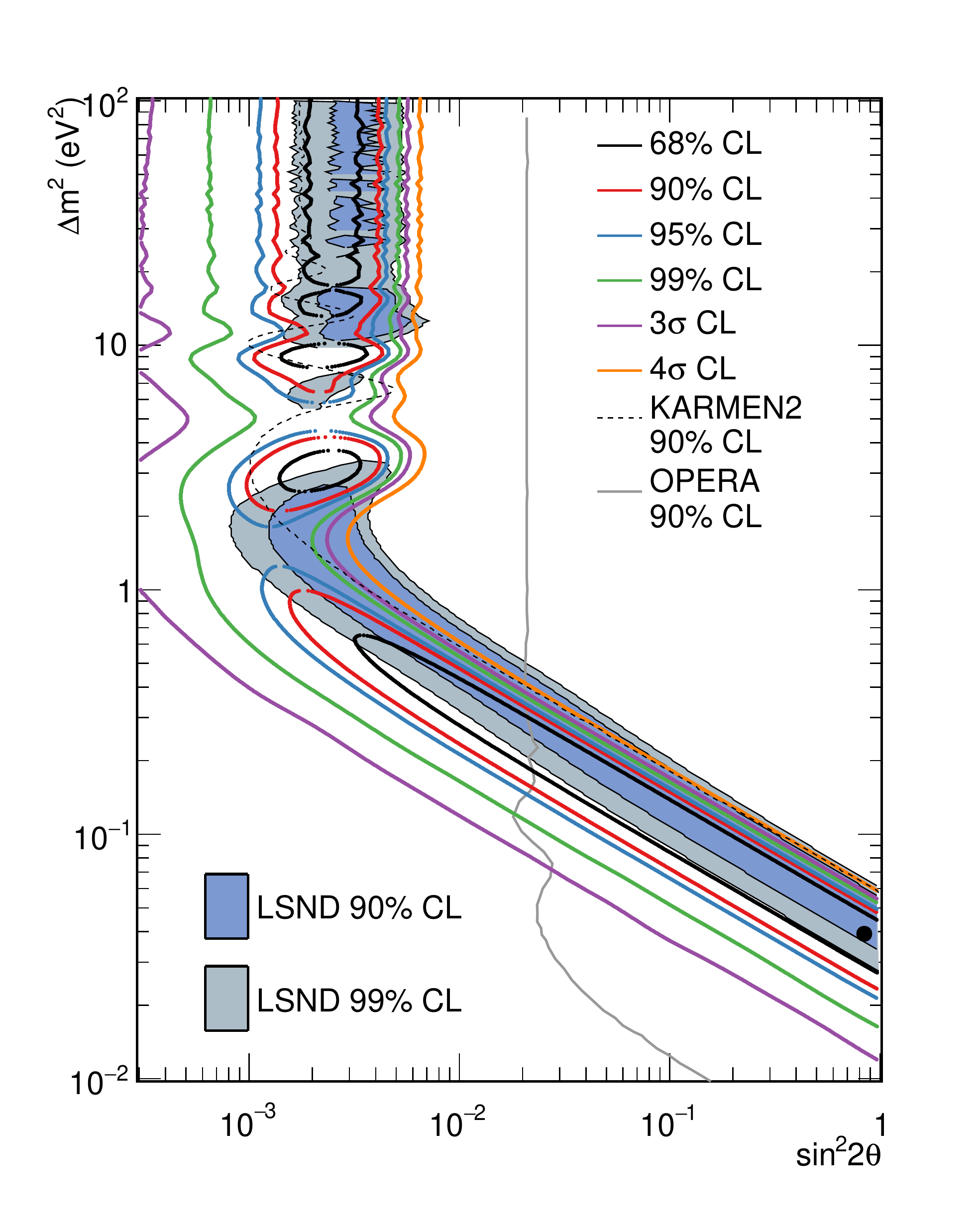}}
 \vspace{-0.3in}
\caption{MiniBooNE allowed regions in neutrino mode  
($12.84 \times 10^{20}$ POT) for events with
$200 < E^{QE}_{\nu} < 3000$ MeV within a two-neutrino oscillation model. 
The shaded areas show the 90\% and 99\% C.L. LSND 
$\bar{\nu}_{\mu}\rightarrow\bar{\nu}_e$ allowed 
regions. The black point shows the MiniBooNE best fit point. 
Also shown are 90\% C.L. limits
from the KARMEN \cite{karmen}
and OPERA \cite{opera} experiments.}
\label{limitab}
\vspace{0.1in}
\end{figure}

A two-neutrino model is assumed for the MiniBooNE oscillation fits in order to compare
with the LSND data.
However, the appearance neutrino experiments appear to be incompatible with the
disappearance neutrino experiments in a 3+1 model \cite{giunti2,kopp2},
and other models \cite{resonant,lorentz,sterile_decay,NSI,sterile_extra} may provide
better fits to the data.
The oscillation parameters are extracted from a combined fit of 
the observed $E_\nu^{QE}$ event distributions for muonlike and electronlike events
using the full covariance matrix described previously in the full energy range 
$200<E_\nu^{QE}<3000$~MeV.
The fit assumes the same oscillation probability for 
both the right-sign $\nu_e$  and wrong-sign  $\bar \nu_e$, and
no $\nu_\mu$, $\bar{\nu}_{\mu}$, $\nu_e$,
or $\bar \nu_e$ disappearance. 
Using a likelihood-ratio technique \cite{mb_oscnew},  the confidence level
values for the fitting statistic, $\Delta\chi^2 =\chi^2(point)-\chi^2(best)$, 
as a function of oscillation parameters, $\Delta m^2$ and $\sin^22\theta$, is determined
from frequentist, fake data studies. The fake data studies also determine the effective 
number of degrees of freedom and probabilities. 
With this technique, the best neutrino oscillation fit in neutrino mode occurs at
($\Delta m^2$, $\sin^22\theta$) $=$ (0.039 eV$^2$, 0.84),
as shown in Fig. ~\ref{limitab}. The $\chi^2/ndf$ 
for the best-fit point in the energy range $200<E_\nu^{QE}<1250$~MeV
is 9.9/6.7 with a probability of 15.5\%.
The background-only fit has a $\chi^2$ probability of
0.06\% relative to the best oscillation fit and a $\chi^2/ndf = 24.9/8.7$
with a probability of 0.21\%.
Fig.~\ref{limitab} shows the MiniBooNE closed 
confidence level (C.L.) contours for
$\nu_e$ appearance oscillations in 
neutrino mode in the
$200<E_\nu^{QE}<3000$~MeV energy range. 

Nuclear effects associated with neutrino interactions 
on carbon can affect the reconstruction of the
neutrino energy, $E_\nu^{QE}$, and the determination 
of the neutrino oscillation parameters \cite{Martinietal}.
These effects were studied previously \cite{mb_oscnew,MartiniGiunti} and were found to not affect 
substantially the oscillation fit. In addition, they do not affect the
gamma background, which is determined from
direct measurements of NC $\pi^0$ and dirt backgrounds. 

\begin{figure}[tbp]
\vspace{-0.25in}
 \centerline{\includegraphics[width=9.0cm]{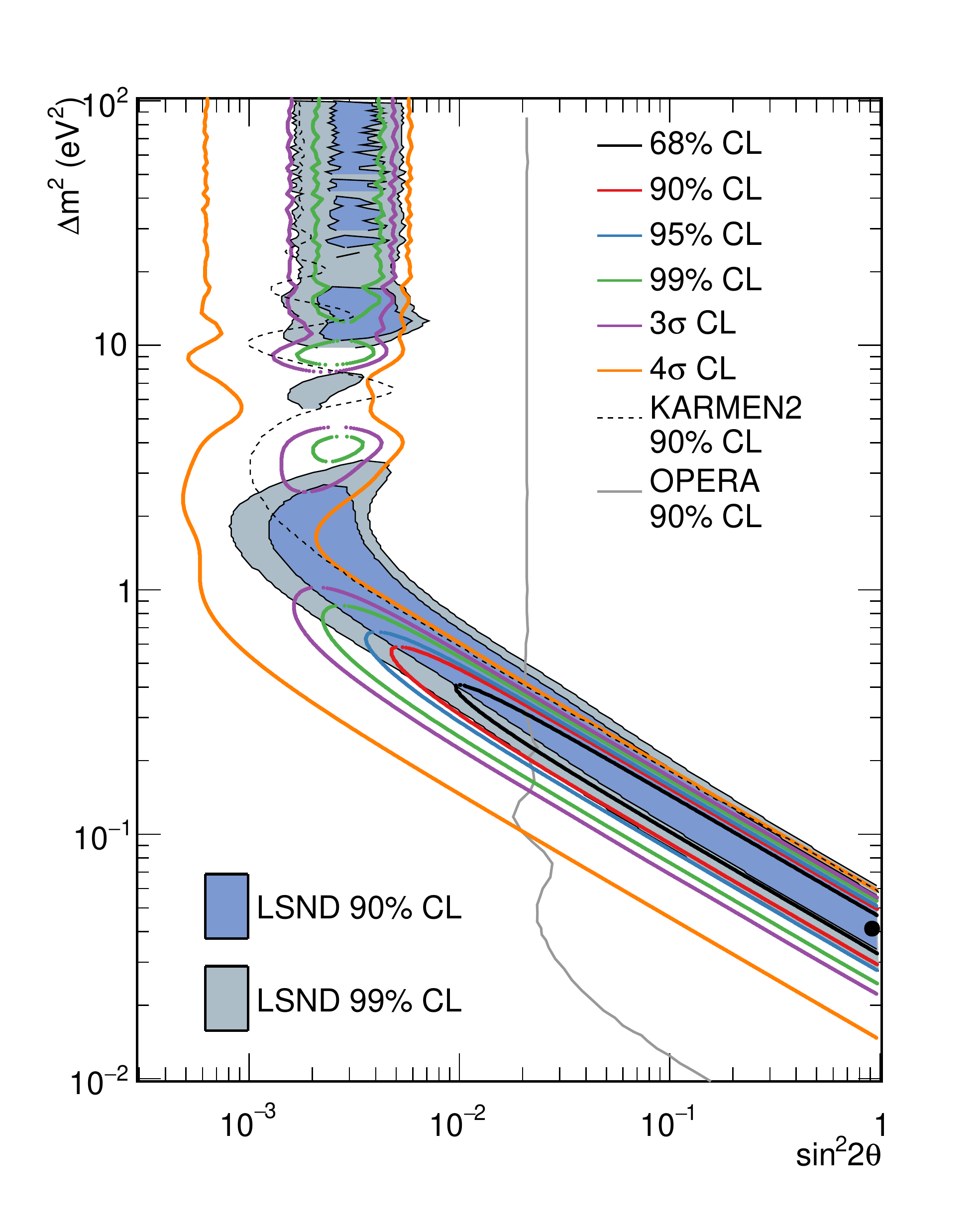}}
 \vspace{-0.3in}
\caption{MiniBooNE allowed regions for a combined neutrino mode ($12.84
  \times 10^{20}$ POT) and antineutrino mode ($11.27 \times 10^{20}$ POT) 
data sets for events with
$200 < E^{QE}_{\nu} < 3000$ MeV within a two-neutrino oscillation model. 
The shaded areas show the 90\% and 99\% C.L. LSND 
$\bar{\nu}_{\mu}\rightarrow\bar{\nu}_e$ allowed 
regions. The black point shows the MiniBooNE best fit point. Also shown are 90\% C.L. limits
from the KARMEN \cite{karmen}
and OPERA \cite{opera} experiments.} 
\label{limitab2}
\vspace{0.1in}
\end{figure}

Fig.~\ref{limitab2} shows the MiniBooNE allowed regions in both neutrino mode and antineutrino 
mode \cite{mb_oscnew}
for events with $200 < E^{QE}_{\nu} < 3000$ MeV within a two-neutrino oscillation model.
For this oscillation fit the entire data set is used and includes the $12.84 \times 10^{20}$ POT 
data in neutrino mode and the $11.27 \times 10^{20}$ POT data in antineutrino mode.
As shown in the figure, the MiniBooNE $1 \sigma$ allowed region lies mostly within the LSND 90\% 
C.L. band, which demonstrates good agreement between the LSND and MiniBooNE signals.
Also shown are 90\% C.L. limits from the KARMEN \cite{karmen}
and OPERA \cite{opera} experiments. The KARMEN2 90\% C.L. limits are outside the MiniBooNE 95\% C.L. allowed region,
while the OPERA 90\% C.L. limits disfavor the MiniBooNE allowed region below approximately 0.3 eV$^2$.
The best combined neutrino oscillation fit occurs at 
($\Delta m^2$, $\sin^22\theta$) $=$ (0.041 eV$^2$, 0.92).
The $\chi^2/ndf$ for the best-fit point in the energy range $200<E_\nu^{QE}<1250$~MeV
is 19.4/15.6 with a probability of $21.1\%$, and the background-only fit
has a  $\chi^2$ probability of $6 \times 10^{-7}$ relative to the best oscillation fit 
and a $\chi^2/ndf = 47.1/17.3$ with a probability of $0.02\%$. 

\begin{figure}[tbp]
\vspace{-0.0in}
\centerline{\includegraphics[angle=0, width=9.0cm]{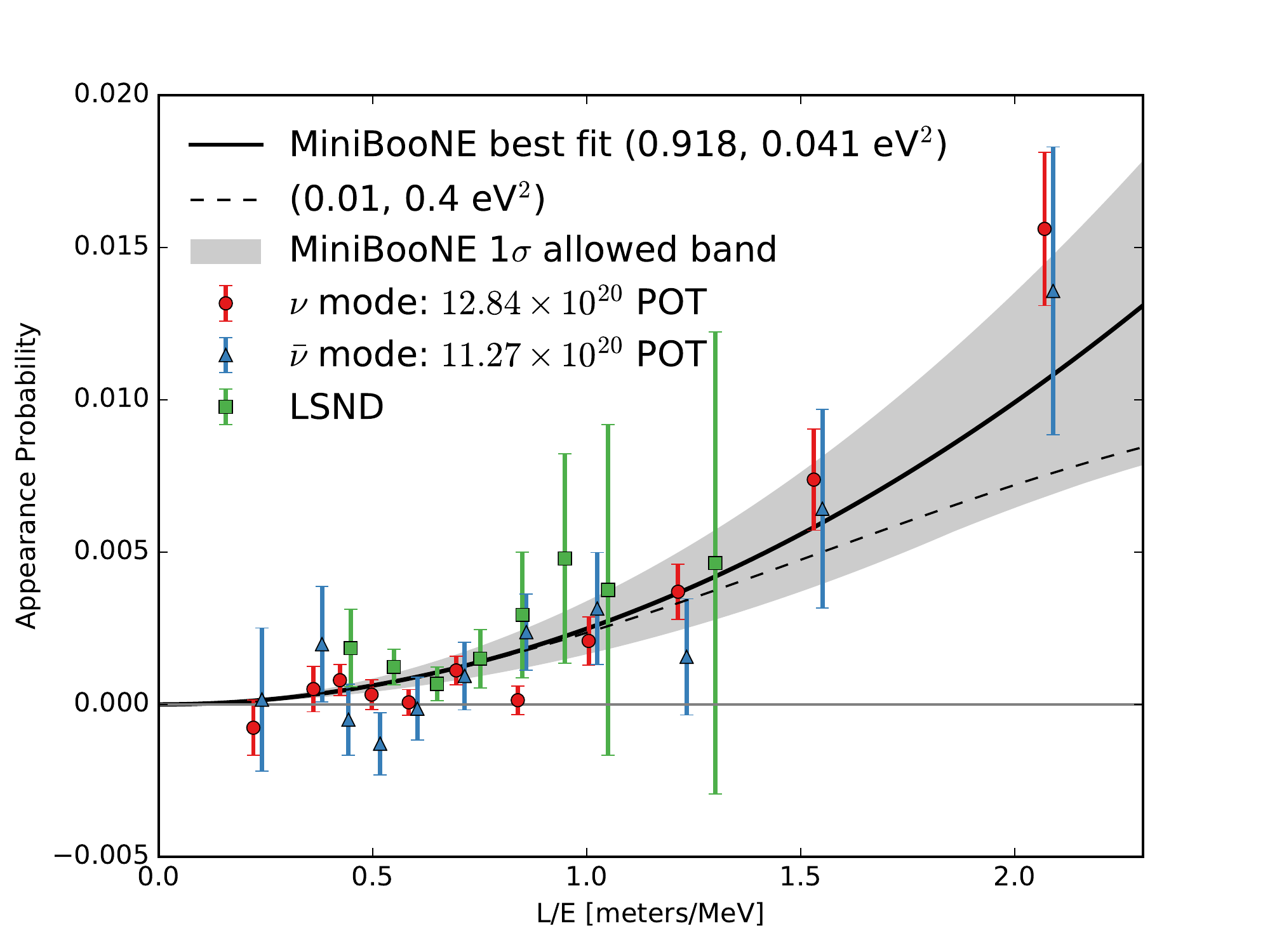}}
\vspace{-0.2in}
\caption{A comparison between the $L/E_\nu^{QE}$
distributions for the MiniBooNE data excesses in neutrino mode
($12.84 \times 10^{20}$ POT) and antineutrino mode ($11.27
\times 10^{20}$ POT) to the $L/E$ distribution from LSND \cite{lsnd}. 
The error bars show
statistical uncertainties only. The curves show fits to the
MiniBooNE data, assuming two-neutrino oscillations, while the shaded area is the MiniBooNE $1 \sigma$ 
allowed band. 
The best-fit curve uses the reconstructed neutrino energy, $E_\nu^{QE}$,
for the MiniBooNE data. The dashed curve shows the example $1 \sigma$ fit point.}
\label{LoverE}
\vspace{0.1in}
\end{figure}

Fig. \ref{LoverE} compares the $L/E_\nu^{QE}$
distributions for the MiniBooNE data excesses in neutrino mode and
antineutrino mode to the $L/E$ distribution from LSND \cite{lsnd}.
The error bars show statistical uncertainties only. As shown in the figure,
there is agreement among all three data sets.
Assuming two-neutrino oscillations, the curves show fits to the
MiniBooNE data described above. 
Fitting both 
MiniBooNE and LSND data, by adding LSND L/E data as additional terms,
the best fit occurs at ($\Delta m^2$, $\sin^22\theta$) $=$ (0.041 eV$^2$, 0.96)
with a $\chi^2/ndf = 22.4/22.4$, corresponding to a probability of $42.5\%$.
The MiniBooNE excess of events in both oscillation probability and L/E spectrum is, therefore, consistent with the
LSND excess of events.
The significance of the combined LSND ($3.8 \sigma$) \cite{lsnd}
and MiniBooNE ($4.7 \sigma$) excesses is $6.0 \sigma$, which is obtained by adding the significances 
in quadrature, as the two experiments have completely different neutrino energies, neutrino fluxes,
reconstructions, backgrounds, and systematic uncertainties.

In summary, the MiniBooNE experiment 
observes a total $\nu_e$ CCQE event excess in both neutrino and antineutrino running modes
of $460.5 \pm 99.0$ events ($4.7 \sigma$) in the 
energy range $200<E_\nu^{QE}<1250$~MeV.  
The MiniBooNE allowed region from a 
two-neutrino oscillation fit to the data, shown in Fig. \ref{limitab2}, 
is consistent with the allowed region reported by the LSND 
experiment \cite{lsnd}. 
On the other hand, a two-neutrino oscillation interpretation of the data would require at least 
four neutrino types and indicate physics beyond the three neutrino paradigm.
The significance of the combined LSND and MiniBooNE excesses is $6.0 \sigma$.
All of the major backgrounds are constrained by in situ 
event measurements, so nonoscillation explanations would need to invoke new anomalous
background processes. 
Although the data are fit with a two-neutrino oscillation model,
other models may provide
better fits to the data. The MiniBooNE event excess will be further studied by the Fermilab 
short-baseline neutrino (SBN) program \cite{sbn}.

\begin{acknowledgments}
We acknowledge the support of Fermilab, the Department of Energy,
and the National Science Foundation, and
we acknowledge Los Alamos National Laboratory for LDRD funding. 
\vspace{-0.01in}
\end{acknowledgments}

\vfill

\newpage

\appendix{\bf Appendix: Background Determination \& Data vs Monte Carlo Comparisons}

\begin{figure}[tbp]
\vspace{-0.0in}
\centerline{\includegraphics[angle=0, width=9.0cm]{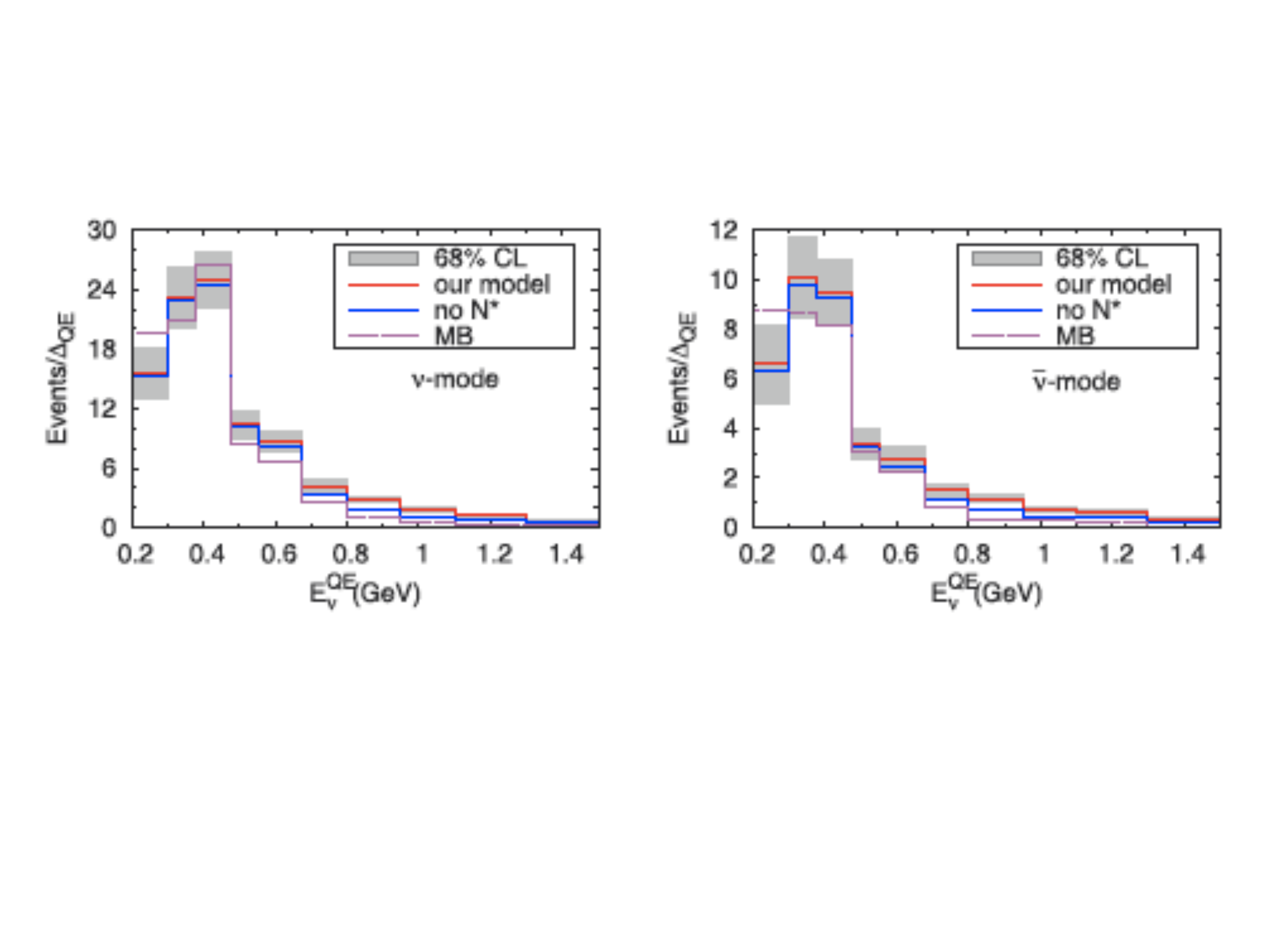}}
\vspace{-0.2in}
\caption{A comparison of the single gamma background estimate from reference \cite{nieves} with that
of MiniBooNE, where good agreement is obtained.}
\label{single_gamma}
\vspace{0.1in}
\end{figure}

Almost all of the backgrounds in the electron-neutrino candidate event sample are determined
directly from MiniBooNE data \cite{ryan}. 
The muon-neutrino charged-current quasi-elastic (CCQE) observed data sample
allows the background determination of both mis-identified $\nu_\mu$ CCQE events from pion decay \cite{mb_numuccqe}
and $\nu_e$ CCQE events from muon decay, as the neutrinos come from the same parent pion and
$\nu_e$ and $\nu_\mu$ cross sections are the same from lepton universality after correcting for
charged lepton mass effects. The $\nu_e$ CCQE background from kaon decay was determined from external measurements and
confirmed by data from the SciBooNE experiment \cite{sciboone_kaon}. In
addition, the neutral current (NC) $\pi^0$ data sample allows the background determination of
both NC $\pi^0$ events \cite{mb_pi0} and single gamma events from $\Delta \rightarrow N \gamma$ decays. Various
theoretical estimates \cite{hill_zhang,nieves} have confirmed the MiniBooNE single gamma background estimate.
Fig. \ref{single_gamma} shows a comparison of the single gamma background estimate from reference \cite{nieves} with that
of MiniBooNE, where good agreement is obtained. Single-gamma backgrounds
from external neutrino interactions (``dirt" backgrounds) are estimated using topological and spatial cuts
to isolate the events whose vertices are near the edge of the detector and
point towards the detector center \cite{mb_lowe}. These estimates have been confirmed by preliminary 
measurements of the absolute event time reconstruction of electron-neutrino candidate events, where
a fit to the dirt event background using timing
agrees within 10\% with the background estimate using topological and spatial cuts.

\begin{figure}[tbp]
\vspace{-0.0in}
\centerline{\includegraphics[angle=0, width=9.0cm]{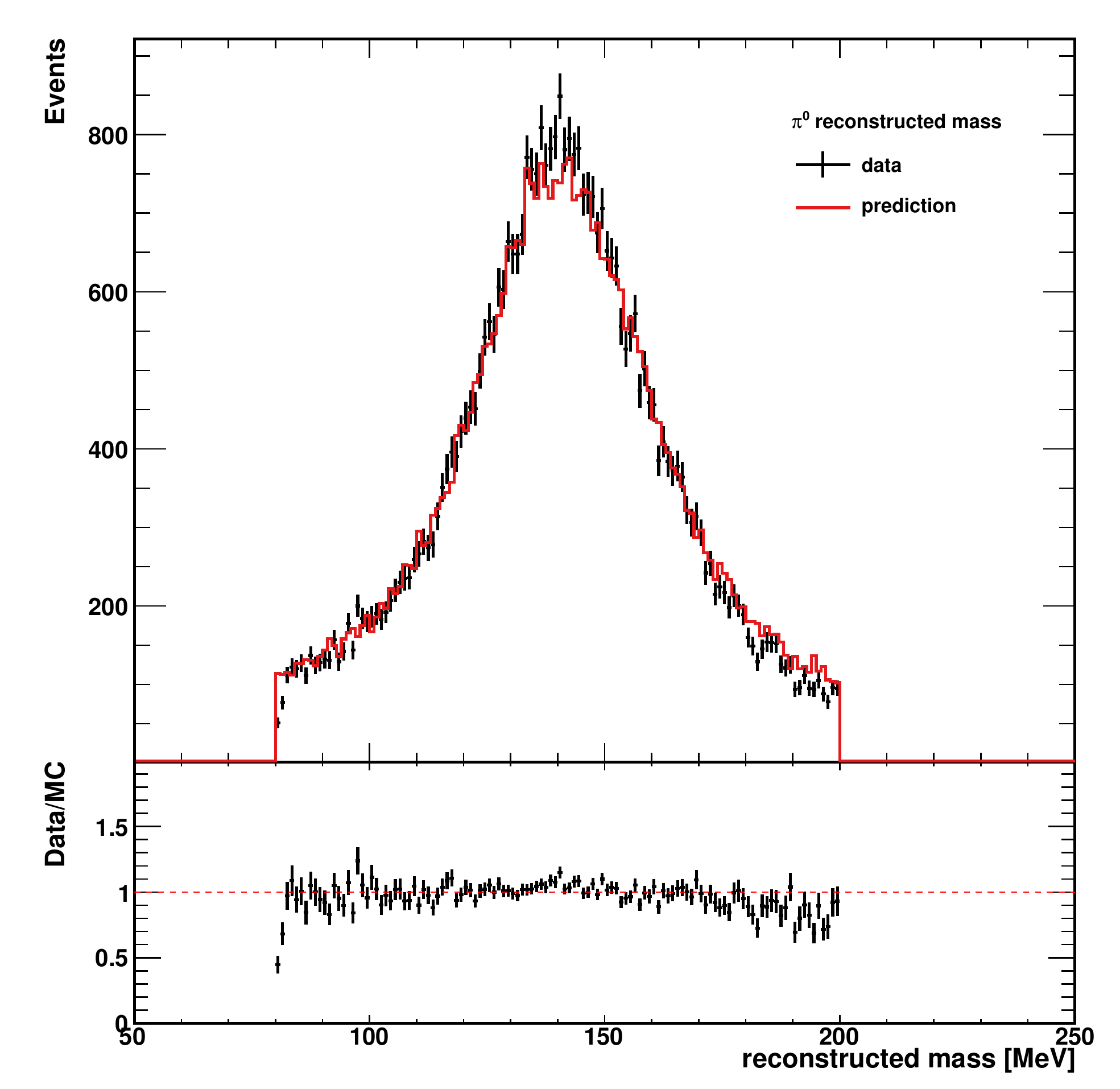}}
\vspace{-0.2in}
\caption{An absolute comparison of the $\pi^0$ reconstructed mass distribution
between the neutrino data ($12.84 \times 10^{20}$ POT)
and the simulation for NC $\pi^0$ events (top). Also shown
is the ratio between the data and Monte Carlo simulation (bottom).
The error bars show only statistical uncertainties.}
\label{pi0mass}
\vspace{0.1in}
\end{figure}

\begin{figure}[tbp]
\vspace{-0.0in}
\centerline{\includegraphics[angle=0, width=9.0cm]{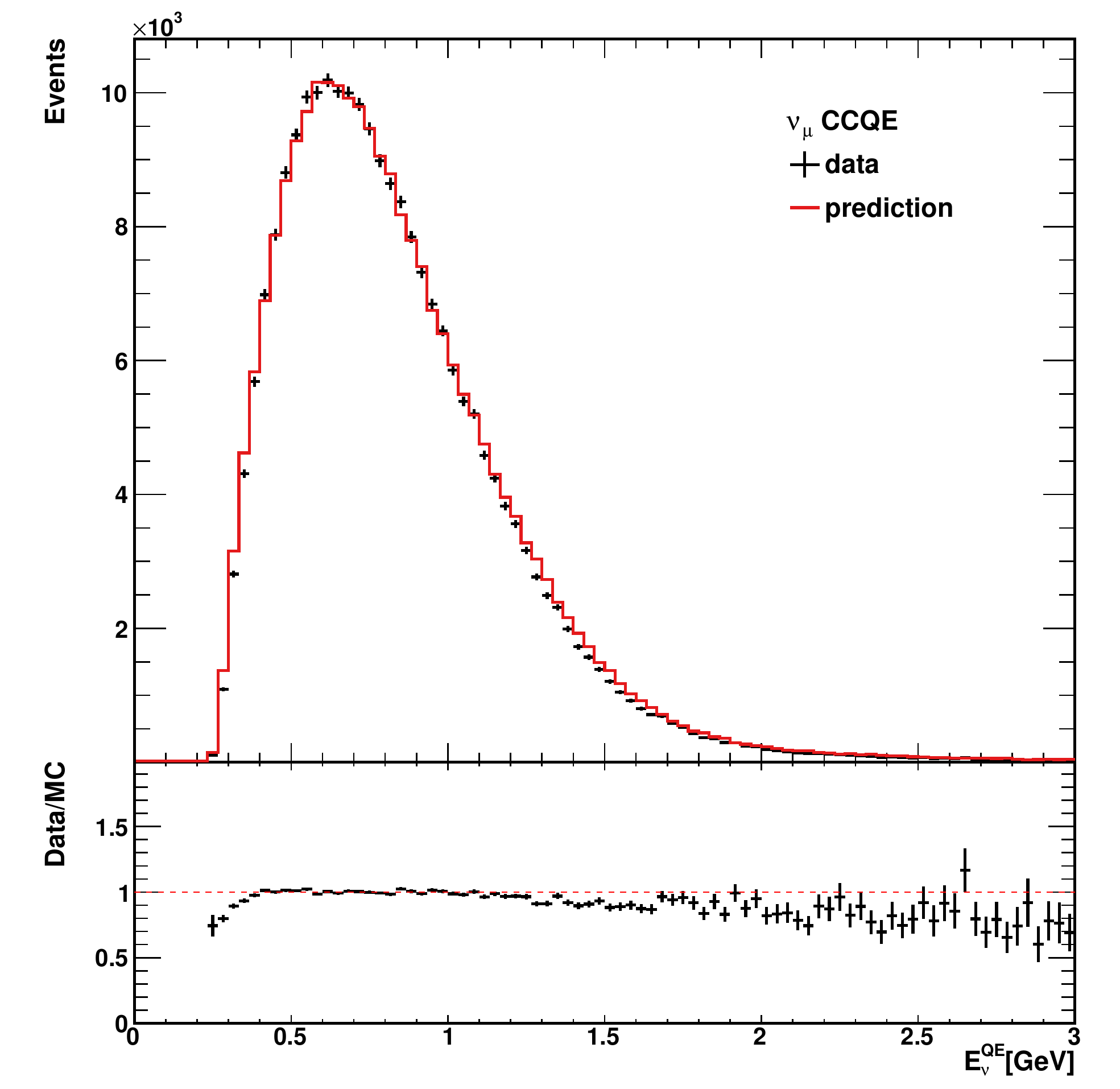}}
\vspace{-0.2in}
\caption{An absolute comparison of the reconstructed neutrino energy distribution
for CCQE events between the neutrino data ($12.84 \times 10^{20}$ POT)
and the simulation (top). Also shown
is the ratio between the data and Monte Carlo simulation (bottom).
The error bars show only statistical uncertainties.}
\label{numuEnuqe}
\vspace{0.1in}
\end{figure}

In order to demonstrate that the MiniBooNE background estimates are reliable, 
various comparisons between the neutrino data, corresponding to $12.84 \times 10^{20}$ 
protons on target (POT), 
and the Monte Carlo simulation have been performed to
check and confirm the accuracy of the simulation. 
Fig. \ref{pi0mass} shows an absolute comparison of
the $\pi^0$ reconstructed mass distribution
between the data and the simulation for NC $\pi^0$ events. Excellent
agreement is obtained, and the ratio of the number of data events (42,483) to the number of
Monte Carlo events (42,530) in the mass range from 80 to 200 MeV/c$^2$
is equal to 0.999. Fig. \ref{numuEnuqe} shows an
absolute comparison of
the reconstructed neutrino energy distribution for CCQE events
between the data and the simulation. Excellent
agreement is also obtained, and the ratio of the number of data events (232,096) to the number of
Monte Carlo events (236,145) is equal to 0.983. 

\begin{figure}[tbp]
\vspace{-0.0in}
\centerline{\includegraphics[angle=0, width=9.0cm]{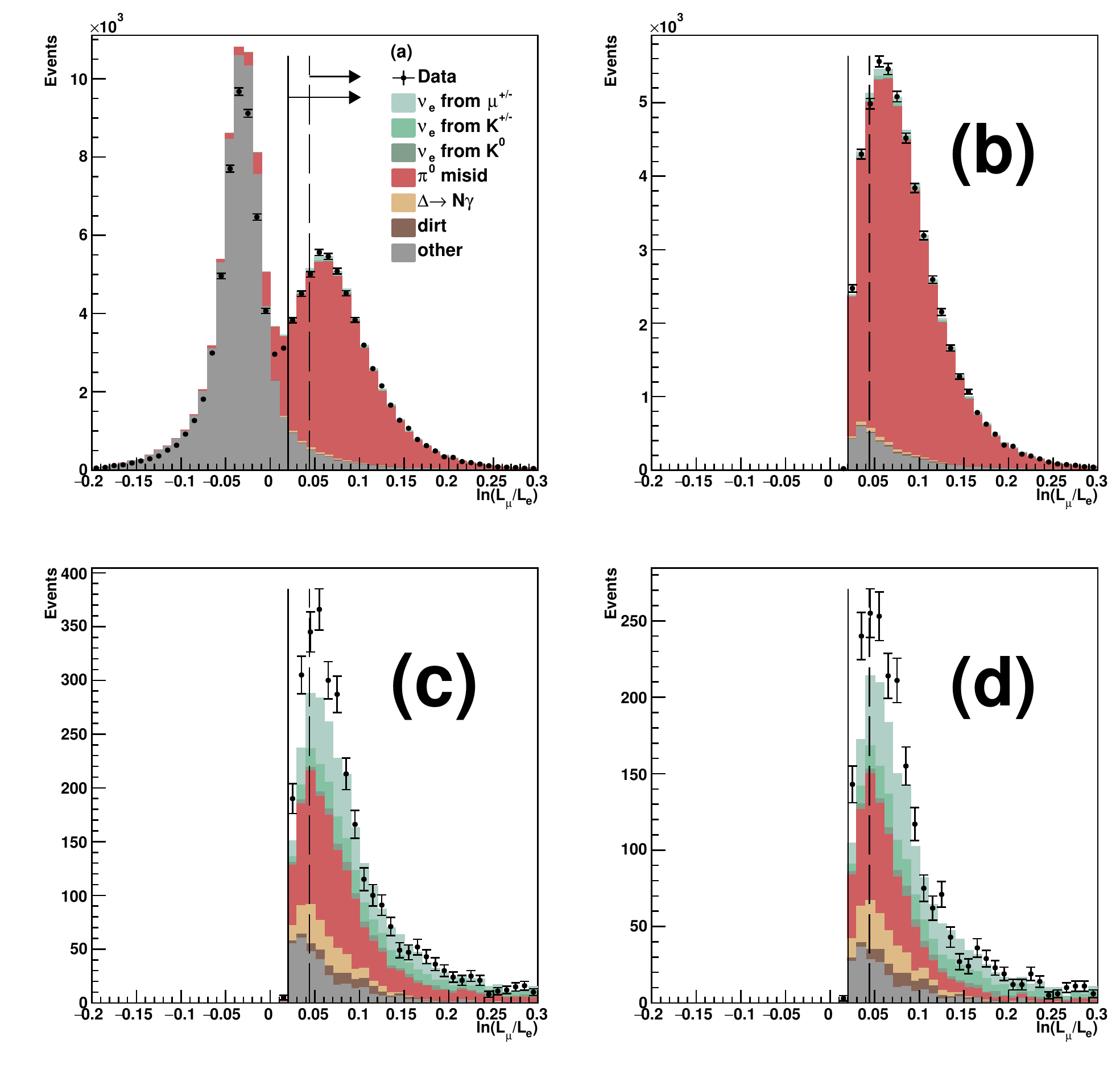}}
\vspace{-0.2in}
\caption{Comparisons between the data and simulation for the electron-muon likelihood distribution
after successive cuts are applied: (a) no PID cut, (b) electron-muon likelihood cut, 
(c) electron-muon plus electron-pion
likelihood cuts, and (d) electron-muon plus electron pion likelihood cuts plus a gamma-gamma mass cut. 
The vertical lines in the figures show the range of energy-dependent 
cut values. The error bars show only statistical uncertainties.}
\label{Lmue}
\vspace{0.1in}
\end{figure}

\begin{figure}[tbp]
\vspace{-0.0in}
\centerline{\includegraphics[angle=0, width=9.0cm]{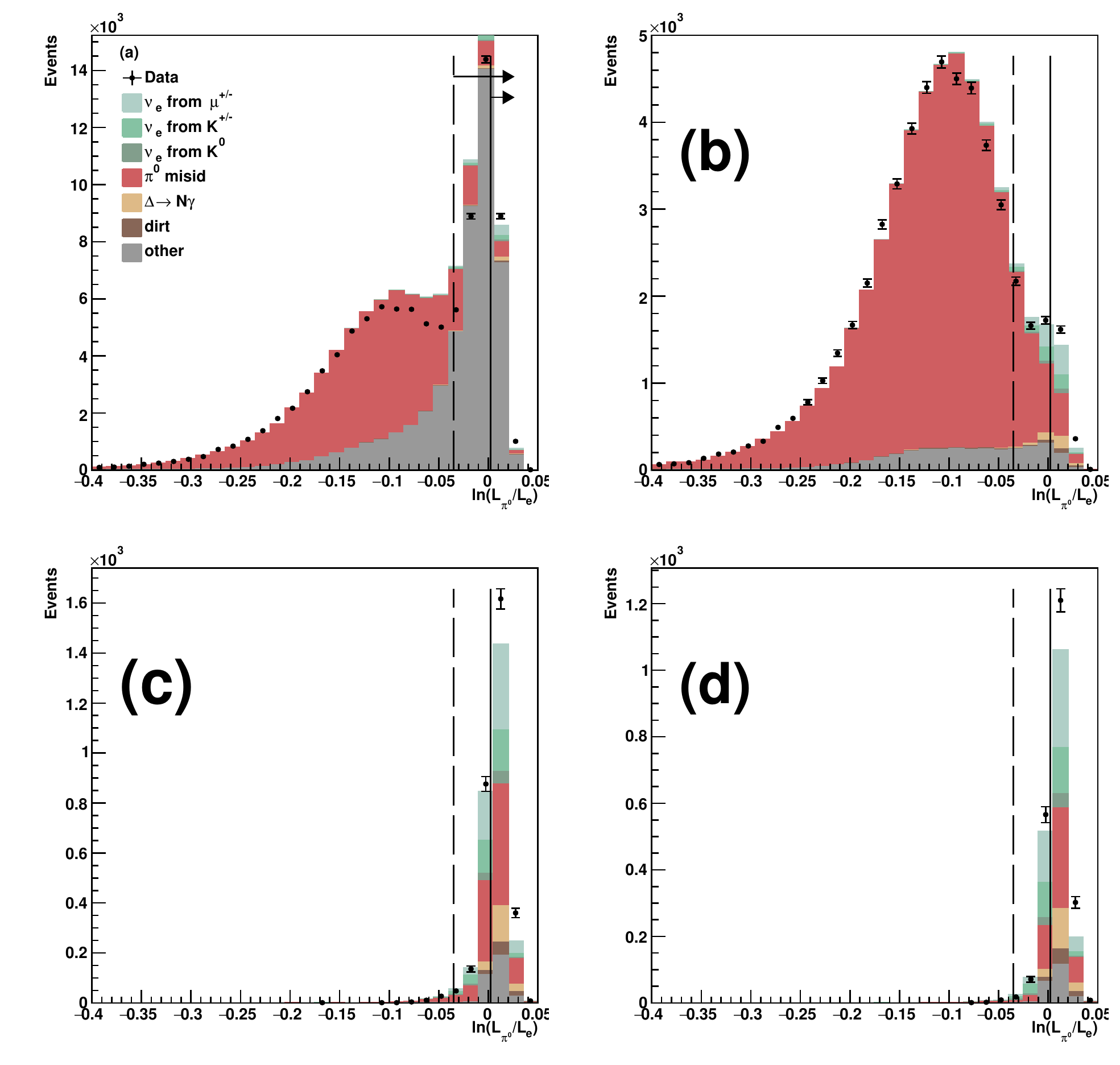}}
\vspace{-0.2in}
\caption{Comparisons between the data and simulation for the electron-pion likelihood distribution
after successive cuts are applied: (a) no PID cut, (b) electron-muon likelihood cut, 
(c) electron-muon plus electron-pion
likelihood cuts, and (d) electron-muon plus electron pion likelihood cuts plus a gamma-gamma mass cut. 
The vertical lines in the figures show the range of energy-dependent
cut values. The error bars show only statistical uncertainties.}
\label{Lpie}
\vspace{0.1in}
\end{figure}

\begin{figure}[tbp]
\vspace{-0.0in}
\centerline{\includegraphics[angle=0, width=9.0cm]{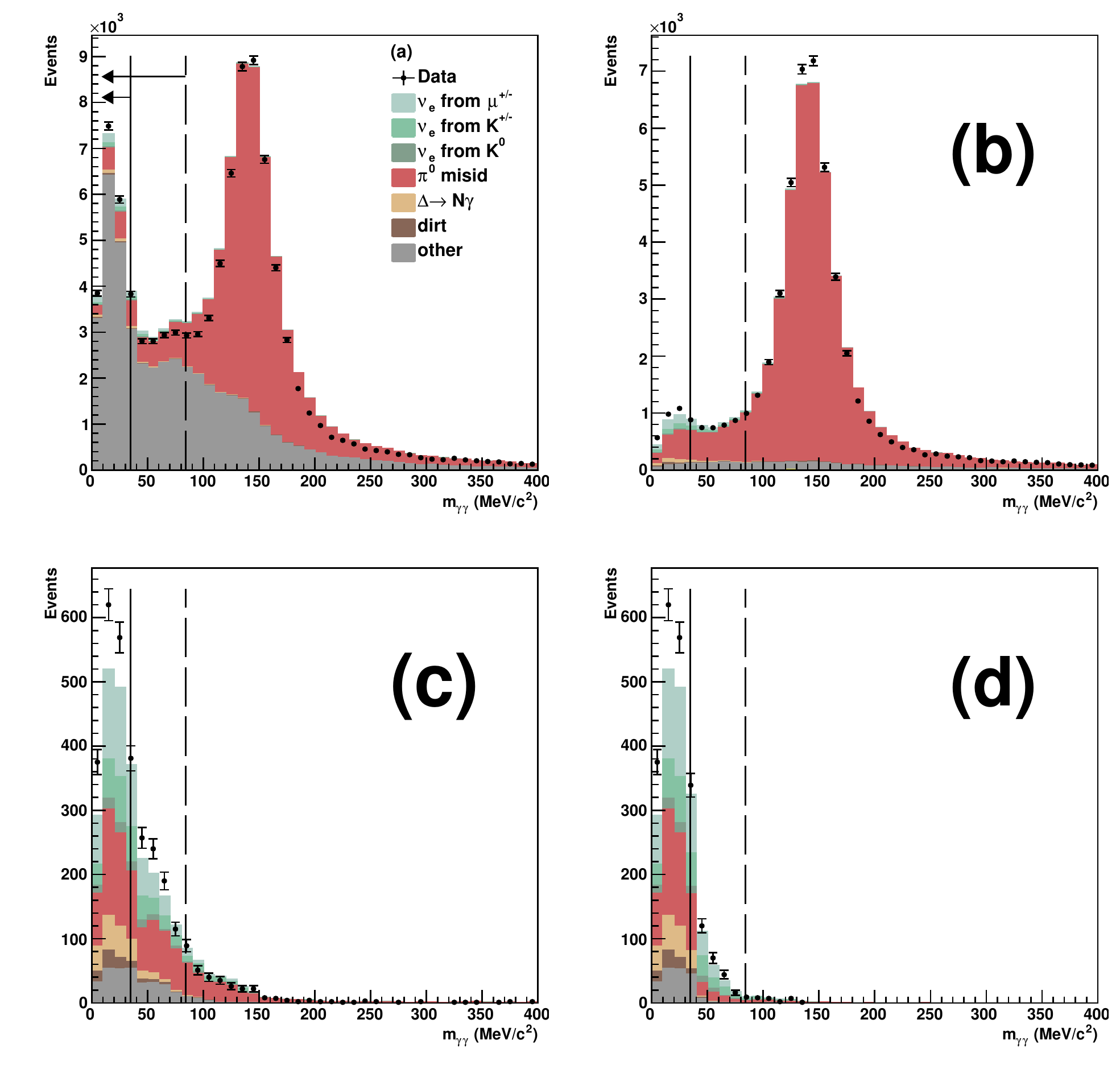}}
\vspace{-0.2in}
\caption{Comparisons between the data and simulation for the gamma-gamma mass distribution
after successive cuts are applied: (a) no PID cut, (b) electron-muon likelihood cut, 
(c) electron-muon plus electron-pion
likelihood cuts, and (d) electron-muon plus electron pion likelihood cuts plus a gamma-gamma mass cut. 
The vertical lines in the figures show the range of energy-dependent
cut values. The error bars show only statistical uncertainties.}
\label{Mgg}
\vspace{0.1in}
\end{figure}

\begin{figure}[tbp]
\vspace{-0.0in}
\centerline{\includegraphics[angle=0, width=9.0cm]{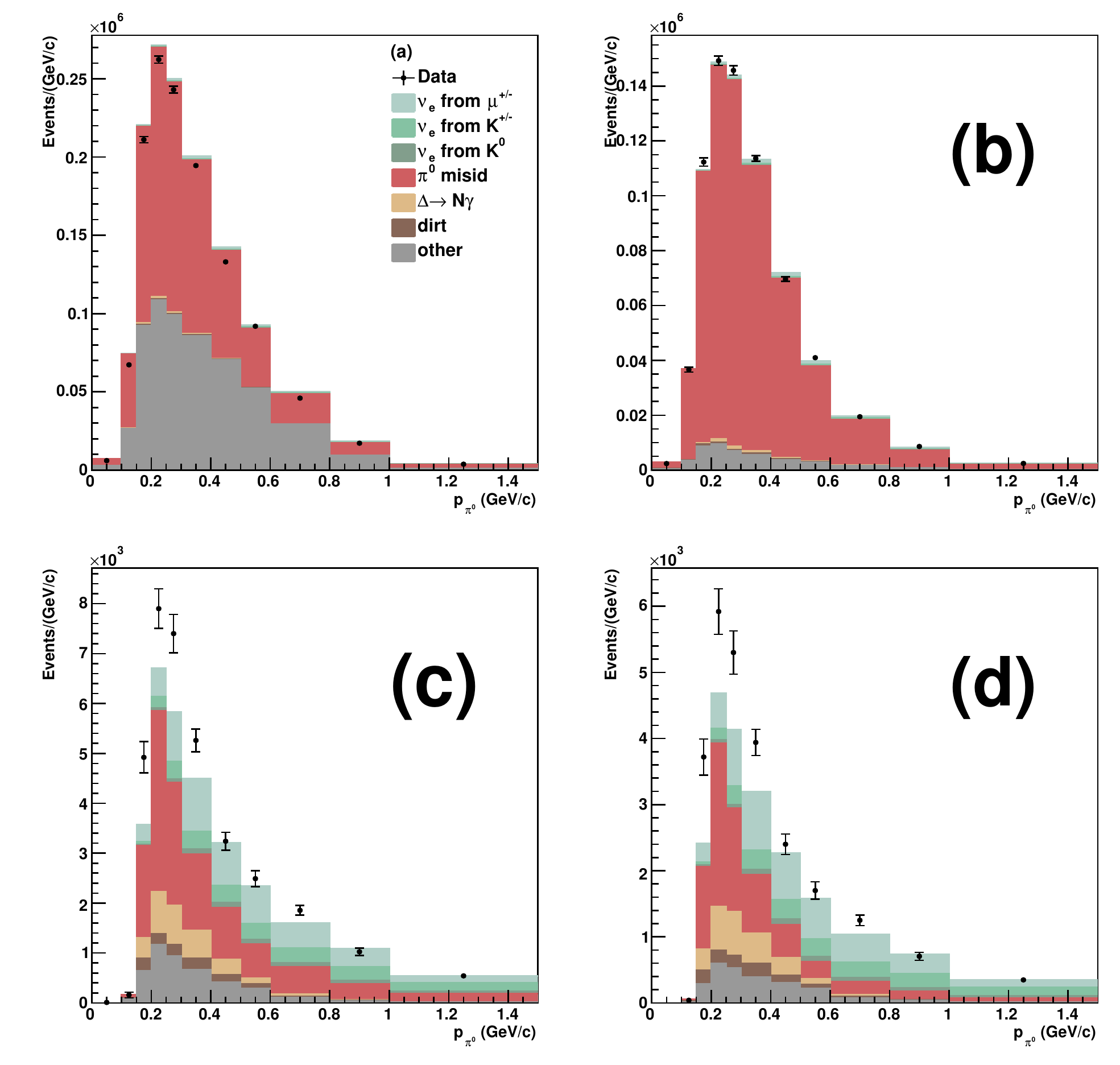}}
\vspace{-0.2in}
\caption{Comparisons between the data and simulation for the momentum distribution
after successive cuts are applied: (a) no PID cut, (b) electron-muon likelihood cut, 
(c) electron-muon plus electron-pion
likelihood cuts, and (d) electron-muon plus electron pion likelihood cuts plus a gamma-gamma mass cut. 
The error bars show only statistical uncertainties.}
\label{Pgg}
\vspace{0.1in}
\end{figure}

\begin{figure}[tbp]
\vspace{-0.0in}
\centerline{\includegraphics[angle=0, width=9.0cm]{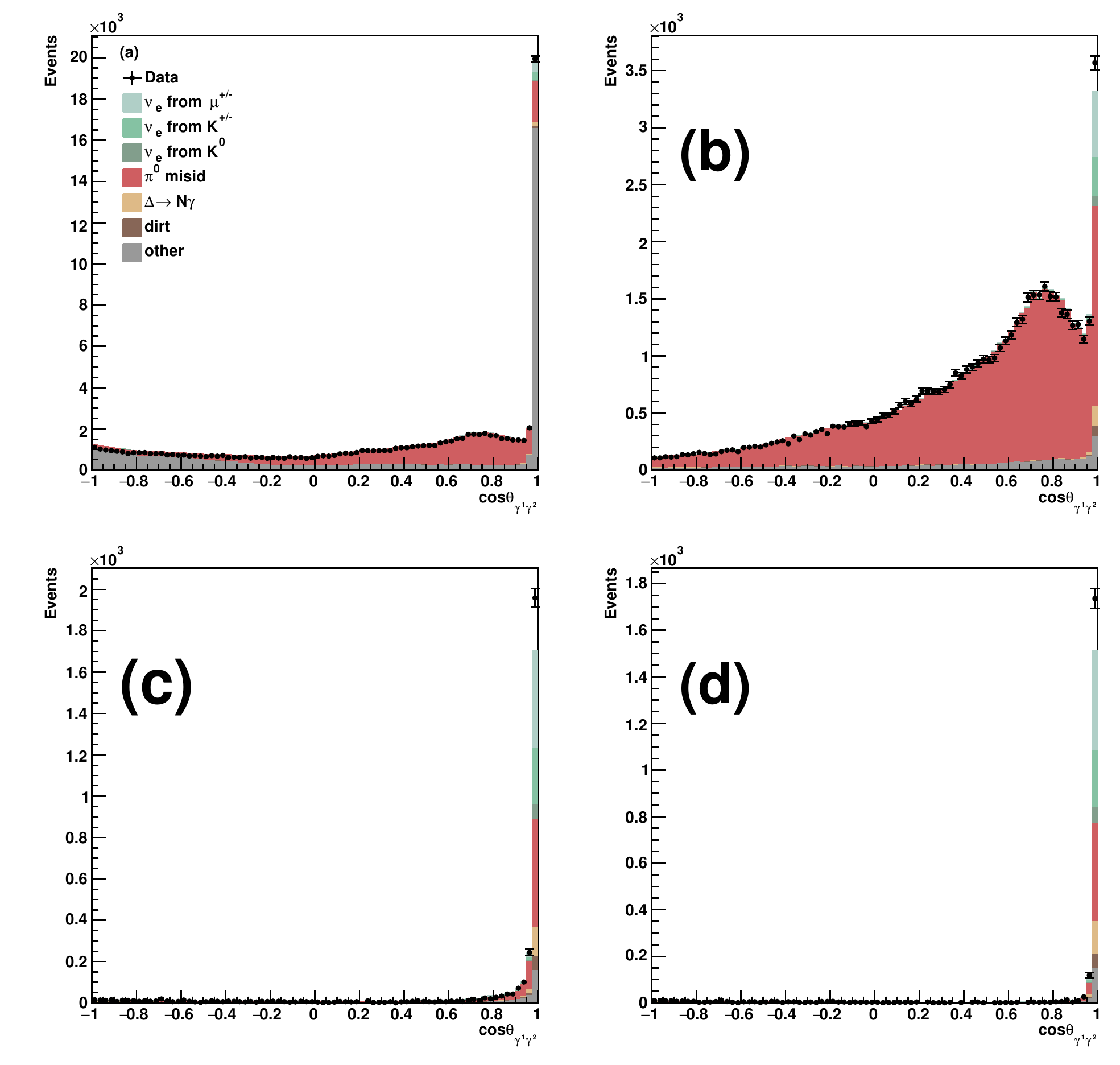}}
\vspace{-0.2in}
\caption{Comparisons between the data and simulation for the gamma-gamma opening angle distribution
after successive cuts are applied: (a) no PID cut, (b) electron-muon likelihood cut,
(c) electron-muon plus electron-pion
likelihood cuts, and (d) electron-muon plus electron pion likelihood cuts plus a gamma-gamma mass cut.
The event excess occurs almost entirely for opening angles less than 13 degrees.
The error bars show only statistical uncertainties.}
\label{opening_angle}
\vspace{0.1in}
\end{figure}

In order to check the particle identification (PID) cuts, 
Figs. \ref{Lmue}, \ref{Lpie}, and \ref{Mgg} show comparisons 
between the data and simulation for the electron-muon likelihood distribution, the electron-pion
likelihood distribution, and the gamma-gamma mass distribution. In each figure, distributions are
shown after successive cuts are applied: no PID cut, electron-muon likelihood cut, electron-muon plus
electron-pion likelihood cuts, and electron-muon plus
electron-pion likelihood cuts and a gamma-gamma mass cut. The last plot in each figure shows distributions with the
final event selection. The vertical lines in the figures show the range of energy-dependent cut values.
Good agreement between the data and the
simulation is obtained outside the cut values, while an excess of events is observed inside the
cut values. Figs. \ref{Pgg} and \ref{opening_angle} show the momentum and gamma-gamma opening angle 
distributions after successive cuts are applied. Good agreement is obtained between the data and Monte
Carlo simulation for the no PID cut distributions, while event excesses are observed after the final
event selection. These five plots also demonstrate that sidebands show good agreement 
between the data and simulation.

\begin{figure}[tbp]
\vspace{-0.0in}
\centerline{\includegraphics[angle=0, width=9.0cm]{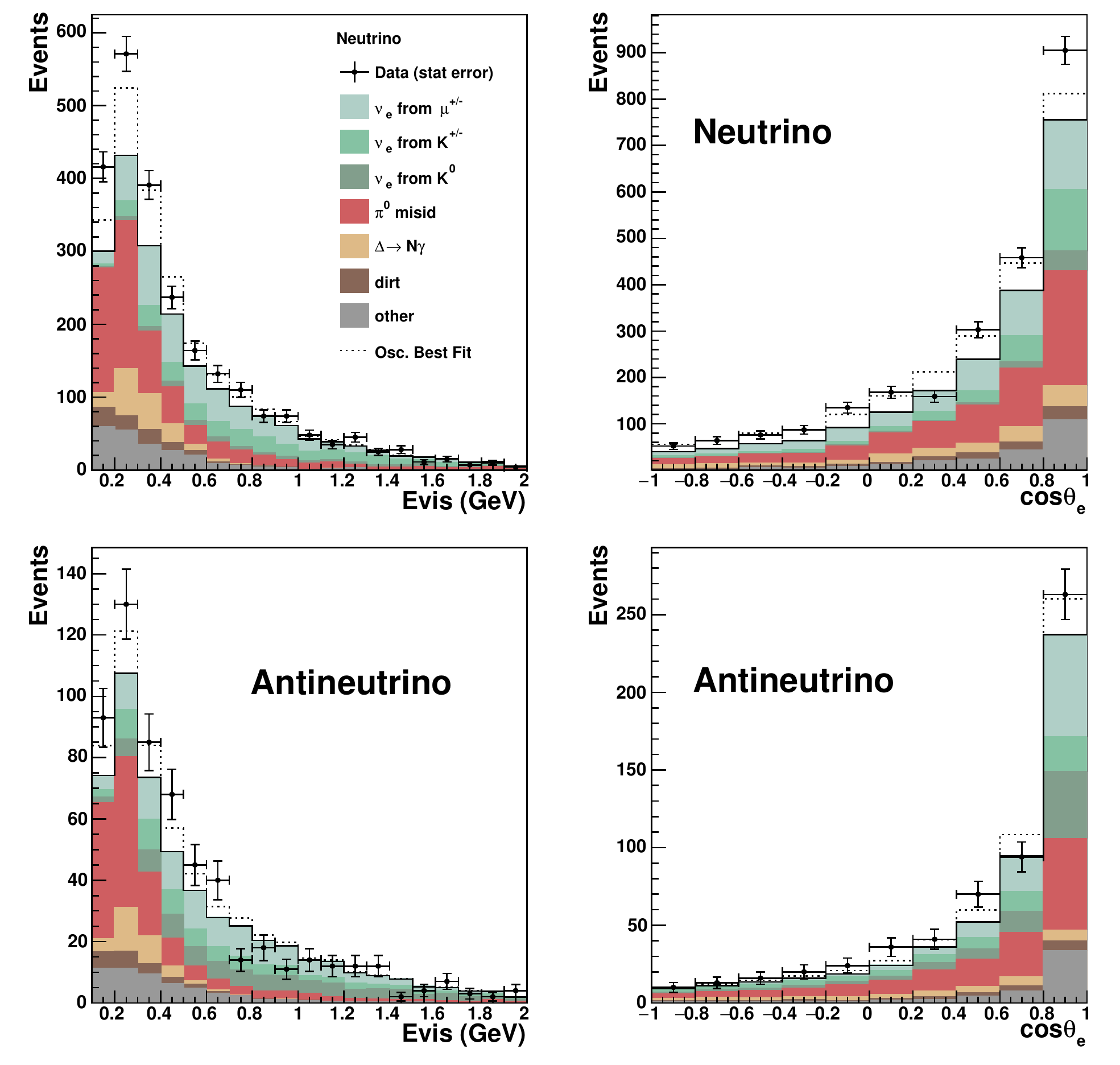}}
\vspace{-0.2in}
\caption{The visible energy (Evis) and $\cos \theta_e$ (Uz)
distributions for the electron-neutrino candidate events in neutrino mode (top)
and antineutrino mode (bottom).
(The error bars show only statistical uncertainties.) Also shown in the
figure are the expectations from all known backgrounds and from the oscillation best fit.}
\label{evis_uz}
\vspace{0.1in}
\end{figure}

\begin{figure}[tbp]
\vspace{-0.0in}
\centerline{\includegraphics[angle=0, width=9.0cm]{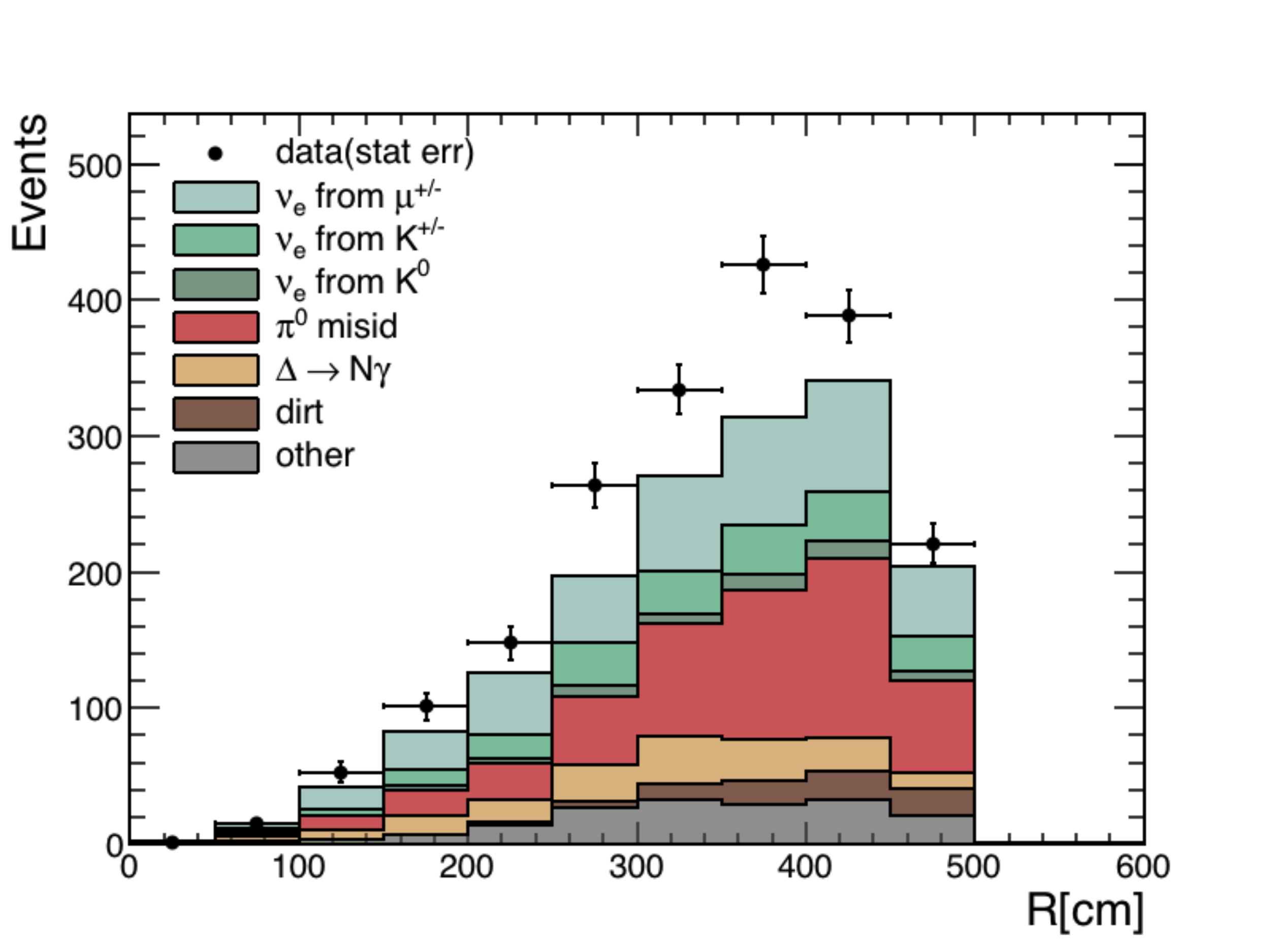}}
\vspace{-0.2in}
\caption{The radius reconstruction in the data is compared to the Monte Carlo simulation. 
The event excess is evenly distributed up to the 5m radius cut.}
\label{Radial}
\vspace{0.1in}
\end{figure}

Fig. \ref{evis_uz} shows the visible energy (Evis) and $\cos \theta_e$ (Uz)
distributions for the electron-neutrino candidate events in neutrino mode (top) and
antineutrino mode (bottom). Also shown in the
figures are the expectations from all known backgrounds and from the oscillation best fit.
These distributions are important because the reconstructed neutrino energy, $E_\nu^{QE}$, is
determined from Evis and Uz. A check on the spatial reconstruction is shown in Fig. \ref{Radial},
where the radius reconstruction in the data is compared to the Monte Carlo simulation. 
As shown in the figure, the event excess is evenly distributed up to the 5m radius
cut. A further check of the electron efficiency was obtained from the reconstruction
of electron events in MiniBooNE that originated from the off-axis NUMI beam (P. Adamson et al.,
Phys. Rev. Lett. {\bf 102}, 211801 (2009)), as the intrinsic $\nu_e$ background was approximately
ten times higher in the NUMI beam than in the BNB. The measured electron rate agreed with the simulation
within errors.

\bigskip

\appendix{\bf Appendix: Stability Checks}

\begin{figure*}[tbp]
\vspace{-0.0in}
\centerline{\includegraphics[angle=0, width=18.0cm]{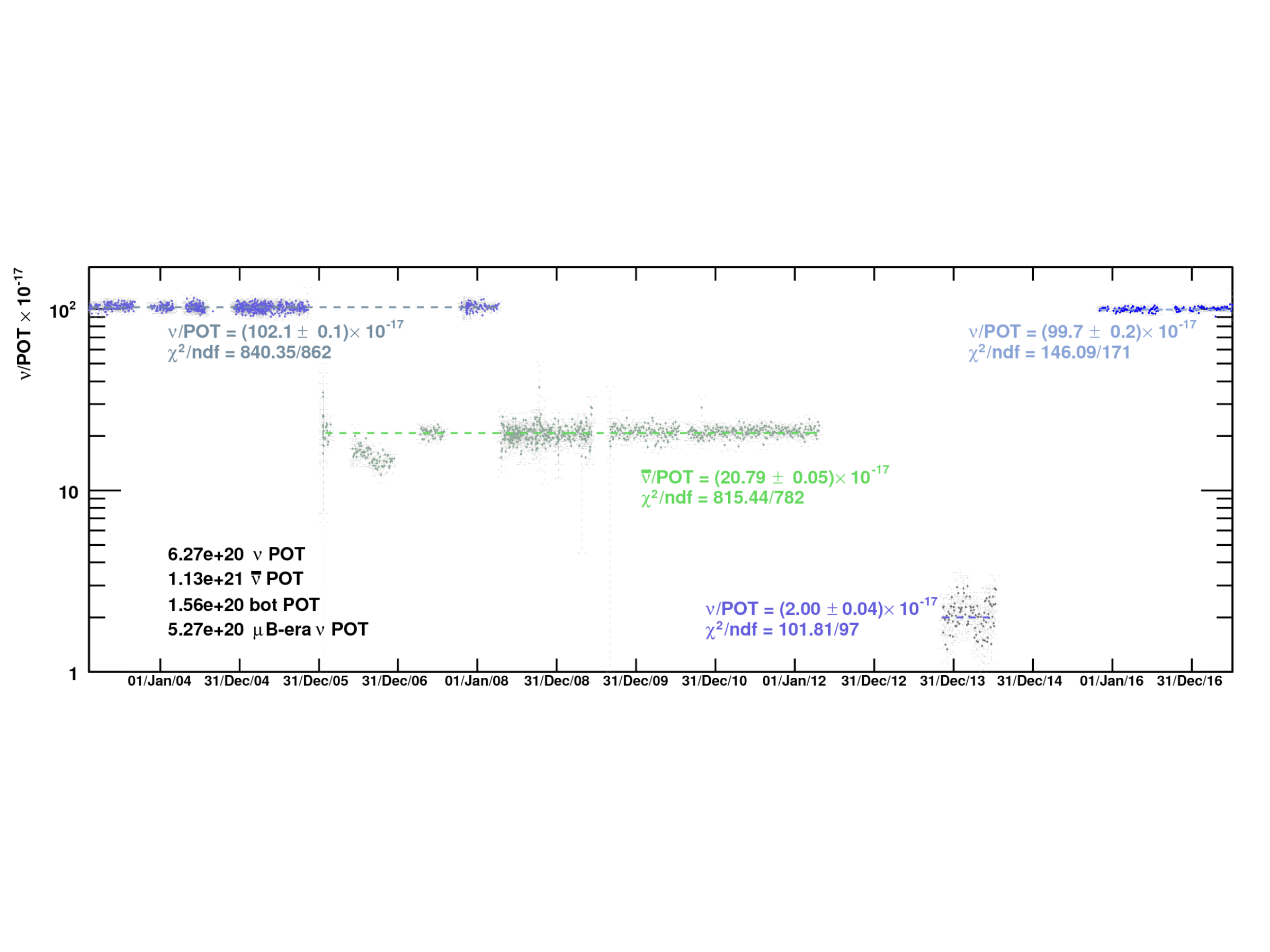}}
\vspace{-0.2in}
\caption{The total number of neutrino events
observed per $10^{17}$ POT over the lifetime of MiniBooNE in neutrino mode, antineutrino mode and beam-dump 
mode.}
\label{events_POT}
\vspace{0.1in}
\end{figure*}

\begin{figure}[tbp]
\vspace{-0.0in}
\centerline{\includegraphics[angle=0, width=9.0cm]{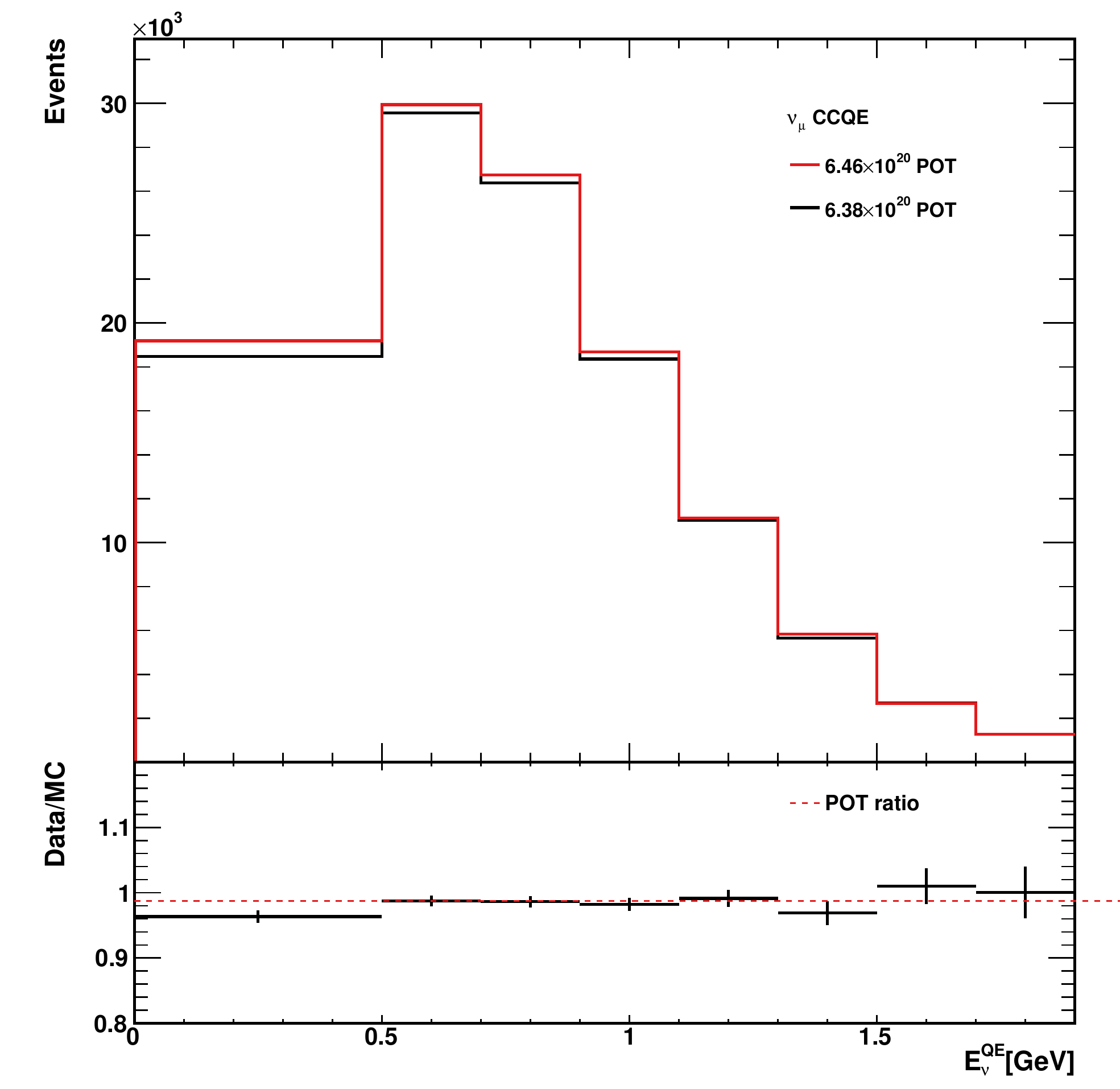}}
\vspace{-0.2in}
\caption{The top plot shows a comparison between the reconstructed $\nu_\mu$ CCQE energy distributions for the
second data set in 2016 and 2017 ($6.38 \times 10^{20}$ POT)
to the first data set ($6.46 \times 10^{20}$ POT). The bottom plot shows the ratio of the second data 
set to the first data set.}
\label{numu_evis}
\vspace{0.1in}
\end{figure}

\begin{figure}[tbp]
\vspace{-0.0in}
\centerline{\includegraphics[angle=0, width=9.0cm]{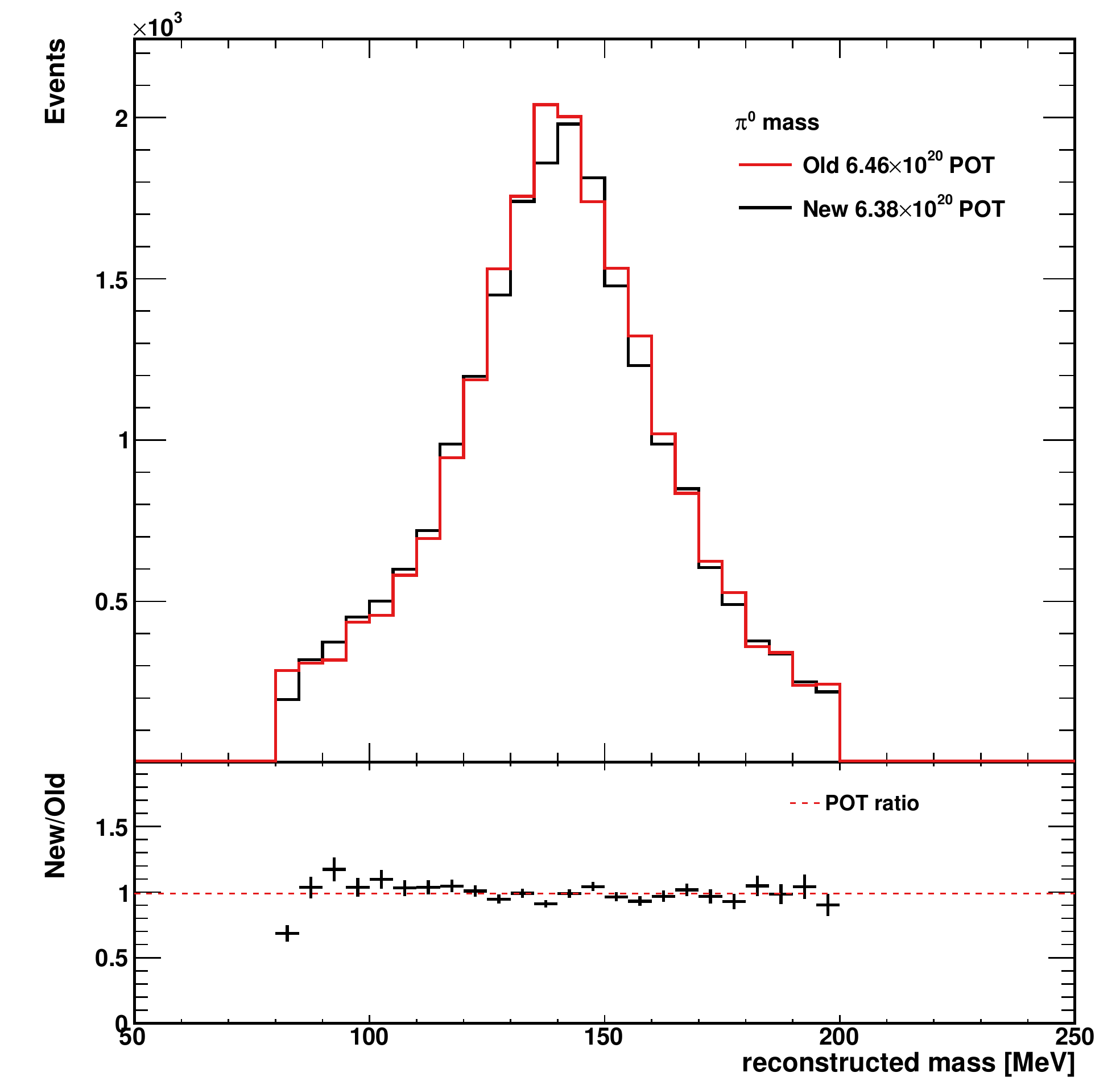}}
\vspace{-0.2in}
\caption{The top plot shows a comparison between the $\pi^0$ mass distributions for the
second data set in 2016 and 2017 ($6.38 \times 10^{20}$ POT) to the first data set
($6.46 \times 10^{20}$ POT). 
The bottom plot shows the ratio of the second data set to the
first data set.}
\label{pi0_mass}
\vspace{0.1in}
\end{figure}

Many checks have been performed on the data, including
beam and detector stability checks. Fig.~\ref{events_POT} shows the total number of neutrino events
observed per $10^{17}$ POT over the lifetime of MiniBooNE in neutrino mode, antineutrino mode and beam-dump
mode. The neutrino mode event
rate of 100 events per $10^{17}$ POT has been stable to $<2\%$ over the 15 year MiniBooNE running period.
This is within the expected errors from 
time variations in BNB performance, such as 
target/horn change, beam rate monitoring, etc.   
A small change in the detector energy response
between the first and second neutrino data set has been
corrected by increasing the measured energy in the second data set by 2\%. 
About half of the energy change 
is from PMT failures in the intervening years, and the remainder
is within the detector response error from gain variations, oil
properties, etc.  With this energy correction, the first and second data sets
are found to agree well. Fig.~\ref{numu_evis}
compares the reconstructed $\nu_\mu$ CCQE energy distributions for the second data set in 2016 and 2017
to the first data set, where good agreement
is obtained. Likewise, Fig.~\ref{pi0_mass} shows
that the $\pi^0$ mass distribution has also not changed.

\bigskip

\appendix{\bf Appendix: Comparing New and Old Neutrino Data}

\begin{figure}[tbp]
\vspace{+0.1in}
\centerline{\includegraphics[angle=0, width=9.0cm]{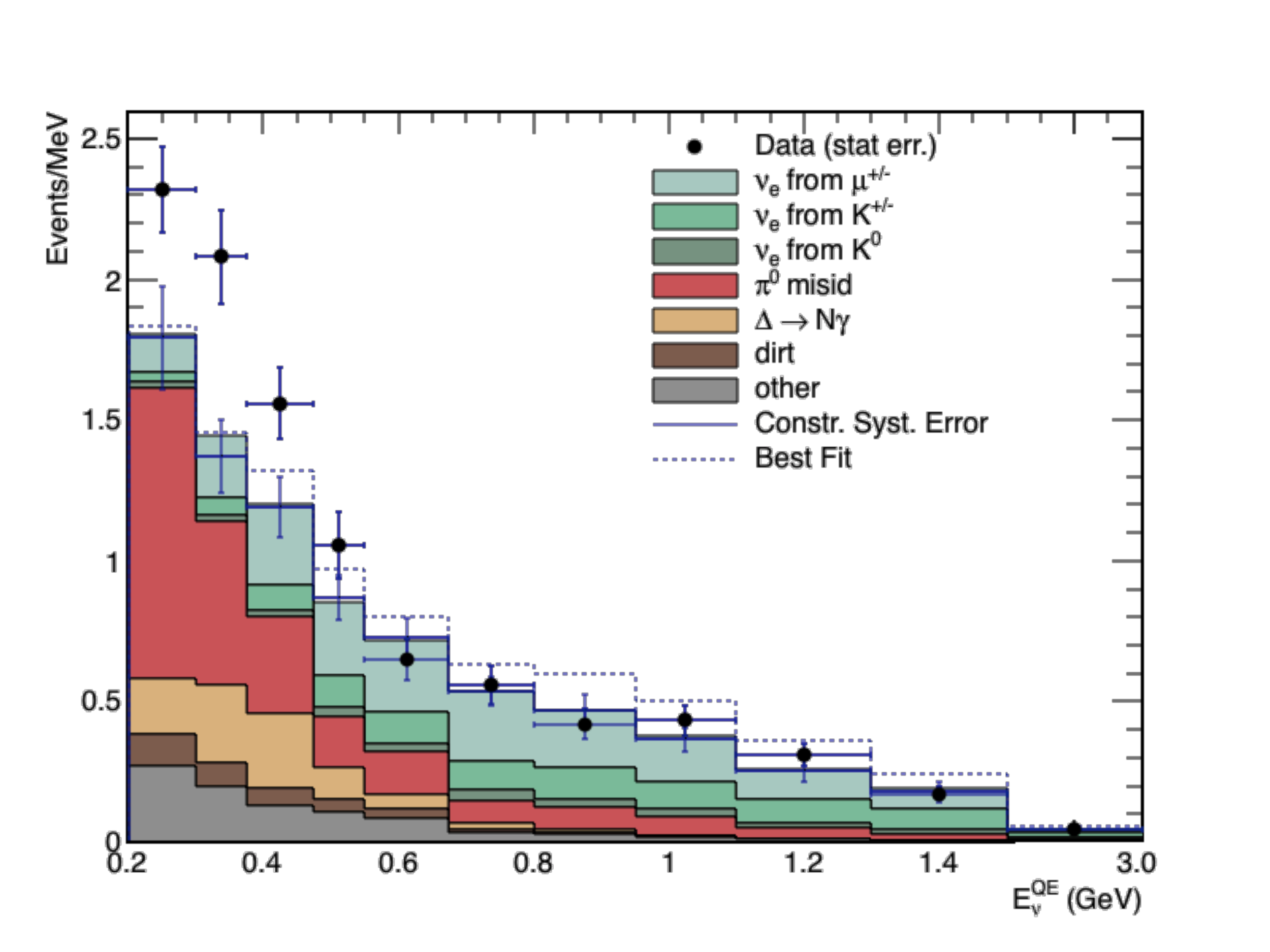}}
\vspace{-0.2in}
\caption{The neutrino mode
$E_\nu^{QE}$ distributions, corresponding to the first $6.46 \times 10^{20}$ POT data set,
for ${\nu}_e$ CCQE data (points with statistical errors) and background (histogram with systematic errors).}
\label{fig6}
\vspace{0.1in}
\end{figure}

\begin{figure}[tbp]
\vspace{+0.1in}
\centerline{\includegraphics[angle=0, width=9.0cm]{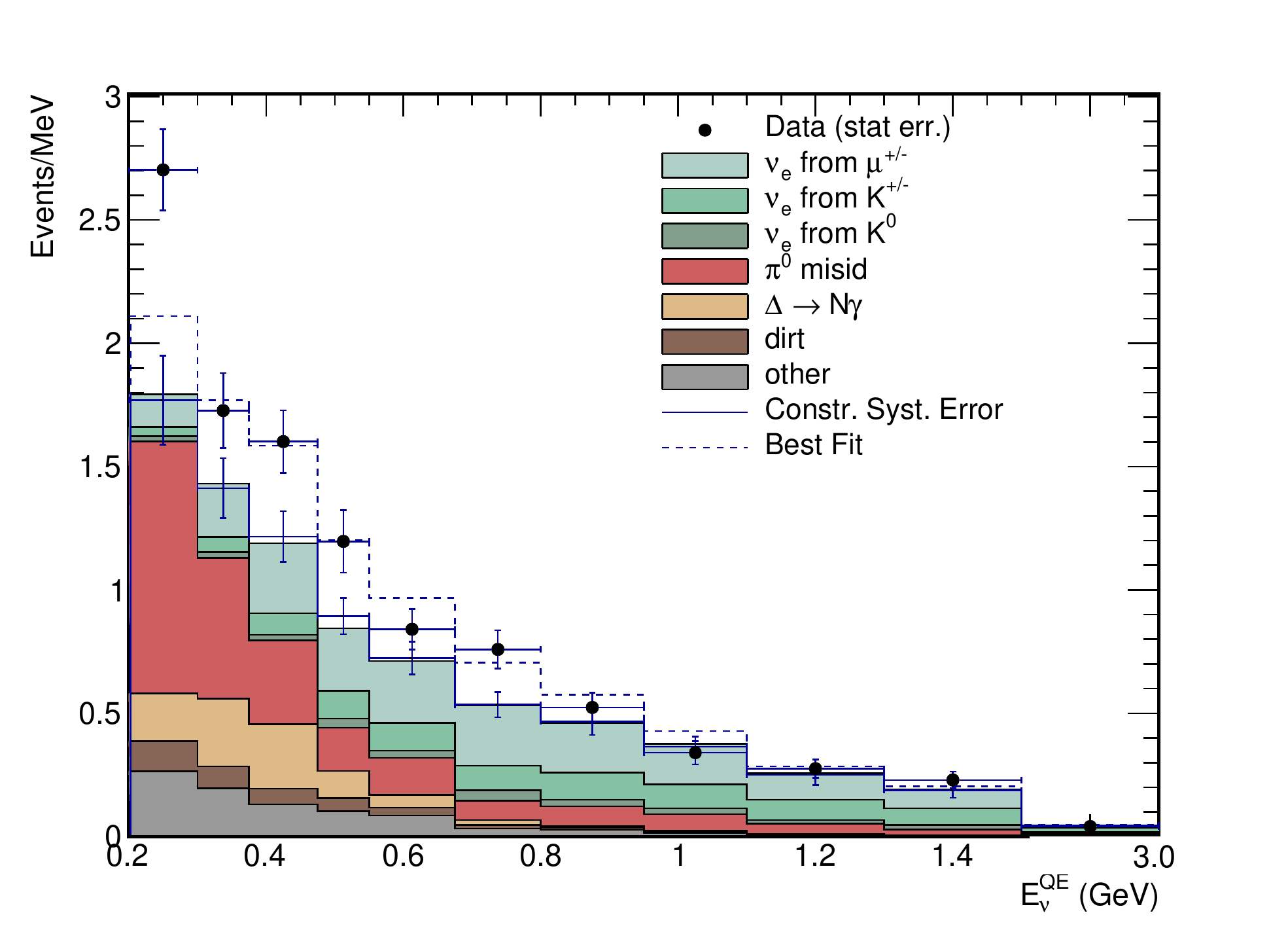}}
\vspace{-0.2in}
\caption{The neutrino mode
$E_\nu^{QE}$ distributions, corresponding to the second $6.38 \times 10^{20}$ POT data set,
for ${\nu}_e$ CCQE data (points with statistical errors) and background (histogram with systematic errors).}
\label{fig8}
\vspace{0.1in}
\end{figure}

\begin{figure}[tbp]
\vspace{+0.1in}
\centerline{\includegraphics[angle=0, width=9.0cm]{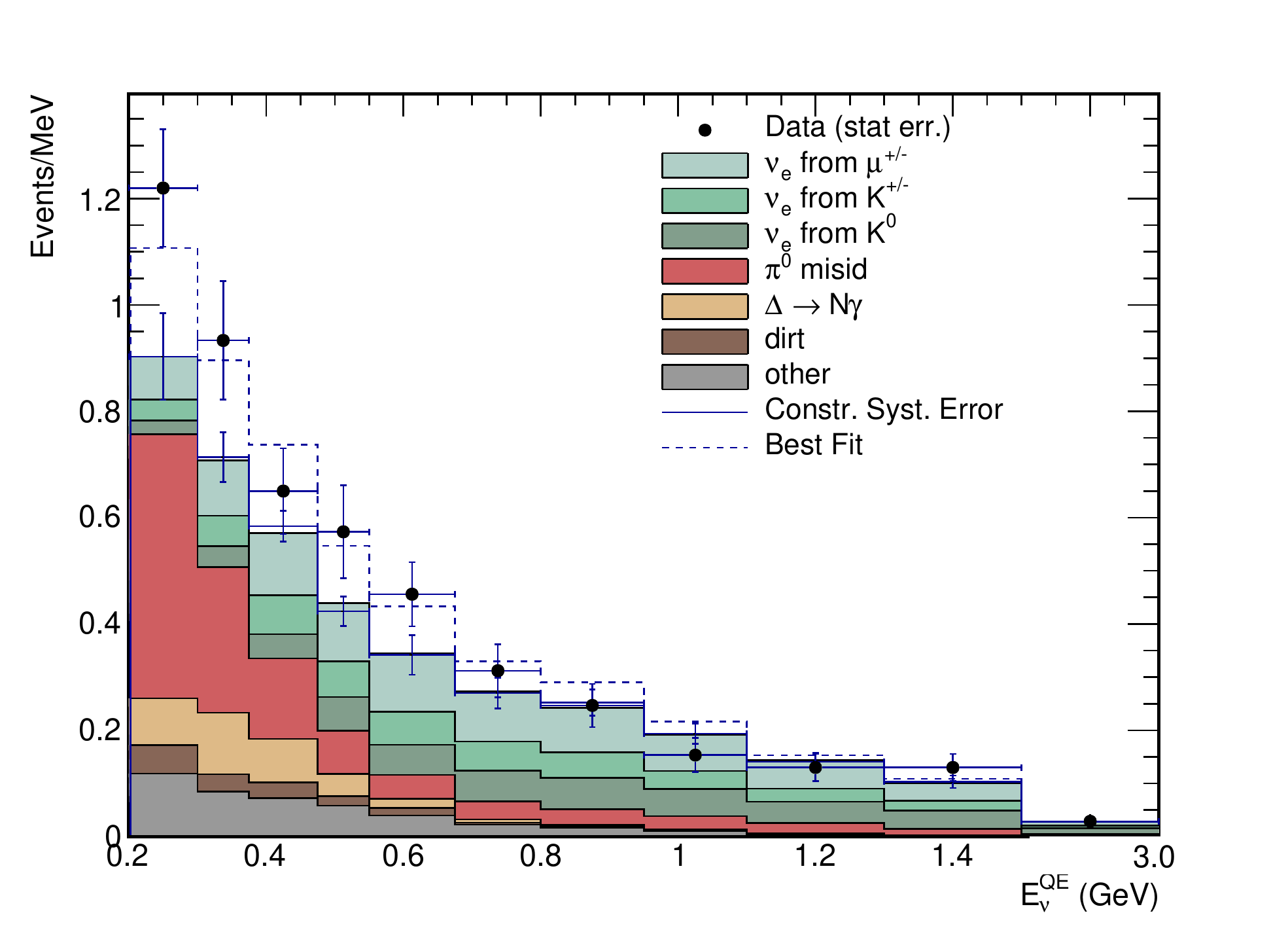}}
\vspace{-0.2in}
\caption{The antineutrino mode
$E_\nu^{QE}$ distributions, corresponding to the published $11.27 \times 10^{20}$ POT data,
for ${\nu}_e$ CCQE data (points with statistical errors) and background (histogram with systematic errors).}
\label{fig10}
\vspace{0.1in}
\end{figure}

Figs. \ref{fig6} and \ref{fig8} show the $E_\nu^{QE}$ distribution for
${\nu}_e$ CCQE data and background in
neutrino mode over the full available energy range for the first $6.46 \times 10^{20}$ POT data set
and the second $6.38 \times 10^{20}$ POT data set.
Fig. \ref{fig10} shows the ${\nu}_e$ CCQE data and background in
antineutrino mode.
Each bin of reconstructed $E_\nu^{QE}$
corresponds to a distribution of ``true'' generated neutrino energies,
which can overlap adjacent bins.
Note that the 162.0 event excess in the $6.46 \times 10^{20}$ POT data
is approximately 1$\sigma$ lower than the average excess, while the
219.2 event excess in the $6.38 \times 10^{20}$ POT data
is approximately 1$\sigma$ higher than the average excess.
In antineutrino mode, a total of 478 data events pass
the $\nu_e$ CCQE event selection requirements with $200<E_\nu^{QE}<1250$~MeV,
compared to a background expectation of $398.7 \pm 20.0 (stat.) \pm 20.5 (syst.) $ events.
The excess is then $79.3 \pm 28.6$ events or a 2.8$ \sigma$ effect.

\vfill

\newpage


\end{document}